\documentclass{aa}  
\usepackage{graphicx}
\usepackage{txfonts,textcomp}
\usepackage{gensymb}

\usepackage{threeparttable}

\DeclareMathOperator{\e}{e}

\usepackage[colorlinks =TRUE, linkcolor = blue,
            urlcolor  = blue,
            citecolor = blue,
            anchorcolor = blue]{hyperref}
\begin{document}

   \title{Disentangling the stellar atmosphere and the focused wind in different accretion states of Cygnus X-1}
   \author{\href{https://orcid.org/0009-0004-1197-5935}{M. Brigitte\thanks{E-mail:maimouna.brigitte@asu.cas.cz}\includegraphics[scale=0.7]{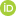}}\inst{1,}\inst{2} \and \href{https://orcid.org/0000-0002-4518-3918}{P. Hadrava\includegraphics[scale=0.7]{ORCID-iD_icon_16x16.png}}\inst{1} \and \href{https://orcid.org/0000-0002-3773-2673}{B. Kubátová\includegraphics[scale=0.7]{ORCID-iD_icon_16x16.png}}\inst{3} \and \href{https://orcid.org/0000-0003-2050-1227}{M. Cabezas\includegraphics[scale=0.7]{ORCID-iD_icon_16x16.png}}\inst{3} \and \href{https://orcid.org/0000-0003-2931-0742}{J. Svoboda\includegraphics[scale=0.7]{ORCID-iD_icon_16x16.png}}\inst{1} \and \href{https://orcid.org/0000-0002-6819-2331}{M. \v{S}lechta\includegraphics[scale=0.7]{ORCID-iD_icon_16x16.png}}\inst{3} \and \href{https://orcid.org/0000-0002-7602-0046}{M. Skarka\includegraphics[scale=0.7]{ORCID-iD_icon_16x16.png}}\inst{3} \and \href{https://orcid.org/0000-0003-0168-9906}{K. Alabarta\includegraphics[scale=0.7]{ORCID-iD_icon_16x16.png}}\inst{4} \and \href{https://orcid.org/0000-0003-1442-4755}{O. Maryeva\includegraphics[scale=0.7]{ORCID-iD_icon_16x16.png}}\inst{3} \and \href{https://orcid.org/0000-0002-3500-631X}{D.M. Russell\includegraphics[scale=0.7]{ORCID-iD_icon_16x16.png}}\inst{4} \and \href{https://orcid.org/0000-0003-1285-4057}{M. C. Baglio\includegraphics[scale=0.7]{ORCID-iD_icon_16x16.png}}\inst{5}} 

   \institute{
   Astronomical Institute of the Czech Academy of Sciences, Bo\v{c}n\'{i} II 1401/1, 14100 Prague 4, Czech Republic
   \and
   Astronomical Institute, Faculty of Mathematics and Physics, Charles University, V Holešovičkách 2, Prague 8, 18000, Czech Republic
    \and
   Astronomical Institute of the Czech Academy of Sciences, Fričova 298, 251 65 Ondřejov, Czech Republic
   \and
   Center for Astrophysics and Space Science (CASS), New York University Abu Dhabi, PO Box 129188, Abu Dhabi, UAE
   \and
   INAF, Osservatorio Astronomico di Brera, Via E. Bianchi 46, I-23807 Merate (LC), Italy}
   
   \date{Received date \ Accepted date}

 
  \abstract
  {In high-mass X-ray binaries (HMXBs), the compact object accretes the strong stellar wind of an O-B supergiant companion star. X-ray flux variations alter the stellar wind's ionization state and optical line profiles, which are important in the determination of the orbital parameters of the system.}
  {We analyzed the state-dependent variability of the line profiles by separating the components coming from the star’s atmosphere and the accreted stream of matter located between the star and the accretion disk (i.e. the focused wind). We then determined the radial velocities and the intensities of the absorption and emission lines with respect to the continuum.}
  {We performed optical high-resolution spectroscopy of the HMXB Cyg X-1 in the hard and soft-intermediate X-ray spectral states, respectively from 2022 and 2023, over multiple orbital phases. We then applied the method of Fourier disentangling to combine the spectra and isolate the stellar atmosphere and focused wind components.}
  {We observe P-Cygni profiles of the $\mathrm{H{\alpha}}$ line in the stellar atmosphere and a wide emission from the focused wind, indicating an outflowing material. While He I $\mathrm{\lambda 6678}$ is in absorption in the stellar atmosphere and not detected in the focused wind, we see a broad emission feature of He II $\mathrm{\lambda 4686}$ in the focused wind. Moreover, we identify an X-ray/optical anticorrelation traced by the strength of the line intensity. The intensity of the lines drops in the soft-intermediate spectral state and the lines are more absorbed at the inferior conjunction of the star.}
  {Our results confirm that the He II emission comes from the focused wind rather than the stellar atmosphere and is produced from the re-scattering of the resonance line due to high-density clumps in the focused wind. The X-ray/optical anticorrelation shows a stronger wind in the low-hard state and the lines are stronger at the inferior conjunction of the star.}


%

    \maketitle
    \section{Introduction}

    Since its discovery in 1964 in X-rays, by an Aerobee rocket \citep{1965AnAp...28..791B}, Cygnus X-1 (Cyg X-1; also known as HDE 226868 or V1357 Cyg for the nomenclature of the companion star) is one of the most studied X-ray binaries. X-ray binaries are systems composed of a compact object accreting matter from a companion star. The conservation of angular momentum leads to the creation of an accretion disk around the compact object; it predominantly emits in X-rays because the matter spiraling inward is heated to millions of kelvins due to viscous dissipation and compressional heating. High-mass X-ray binaries (HMXBs) typically host a massive companion star ejecting a strong stellar wind partially accreted by the compact object. The companion star dominates the optical, ultraviolet (UV), and infrared (IR) emission. Improvements in instrumental precision have enabled a better estimation of the physical parameters of the system. 
    In the 1970s, optical observations of Cyg X-1 from \cite{1971Natur.233..110M}, \cite{1972Natur.235...37W}, and \cite{1972NPhS..240..124B} confirmed it to be a HMXB system with an orbital period of 5.6 days, which has since been revised more precisely to 5.599829(16) days \citep{1999MNRAS.309.1063B, 2011ApJ...742...84O}. A spectroscopic analysis affirmed the presence of an O9.7 blue supergiant companion star \citep{1973ApJ...179L.123W, 1978ApJS...38..309H, 1995A&A...297..556H} and a compact object \citep{1982ApJ...260..240G}, believed to be a black hole \citep{2011ApJ...742...84O}. The system is tightly confined in a quasi-circular orbit with an eccentricity of $\mathrm{0.018 \ \pm \ 0.002}$ and an inclination of $\mathrm{27.1\degree \pm 0.8 \degree}$ \citep{2011ApJ...742...84O, miller-jones_cygnus_2021}. Cyg X-1 is one of the most persistent X-ray sources in our galaxy.  \cite{miller-jones_cygnus_2021} used radio astrometry to update the distance of the system to $\mathrm{2.22^{+ 0.18}_{-0.17}}$ kpc, which implies a higher mass than previously estimated for the black hole and the donor star: $\mathrm{M_{BH} = 21.2 \pm 2.2 \ M_{\odot}}$ and $\mathrm{M_{*} = 40.6 ^{+ 7.7}_{- 7.1} \ M_{\odot}}$, respectively.\\

    X-ray binary systems show variability in their X-ray flux, and some systems transition from a low-hard state to a disk-dominated state -- the high-soft state \citep[e.g.,][]{1998ASPC..138...75C, 1999dicb.conf..245T, 2005A&A...442..555L}. In the low-hard state, the
    soft X-rays coming from the inner accretion disk are up-scattered by hot electrons through Comptonization. The low-density, optically thin hot plasma is often called the "corona" \citep{1975ApJ...195L.101T, 1979Natur.279..506S, 1976ApJ...204..187S}. The geometry of the corona is still debated \citep{2018NatAs...2..652C}. The first X-ray polarimetric measurements of Cyg X-1 via the Imaging X-ray Polarimetry Explorer \citep[IXPE;][]{2022AAS...24024601R} indicate a coronal plasma extended perpendicular to the jet axis and parallel to the accretion disk \citep{2022Sci...378..650K}. In the high-soft state, the X-ray spectrum is dominated by the thermal emission from the disk \citep{Cui:1996ys}. During the accretion state transitions, the X-ray flux increases by several orders of magnitude, directly impacting the stellar wind's ionization state. \\
    \indent For Cyg X-1, the flux increases on average from 0.2 to 2 Crab ($\mathrm{0.48}$ to $\mathrm{4.8 \times 10^{-8} \ erg \ s^{-1} \ cm^{-2}}$) but never completely reaches the thermal-dominant soft state. The source remains in the so-called soft-intermediate state, where the accretion disk emission dominates but the coronal emission is still present.  In this paper we use the term "high-soft-intermediate state" to refer to the softest state of Cyg X-1. The reason for these transitions is still unknown. Previous studies of Cyg X-1 show a correlation between the X-ray emission from the accretion disk and the optical emission from the stellar wind \citep{1999MNRAS.309.1063B, 2003ApJ...583..424G, 2008AJ....136..631Y}. \cite{2003ApJ...583..424G} used the $\mathrm{H{\alpha}}$ line as a proxy for the stellar wind and studied the emission line profile with respect to the X-ray flux between 2001 and 2002. They performed a time interpolation of the Rossi X-ray Timing Explorer (RXTE) data due to the absence of simultaneous observations in the optical and X-rays.  \cite{2008AJ....136..631Y} performed a similar study between 2001 and 2006. Both studies find an anticorrelation between the X-ray flux and the equivalent width of the $\mathrm{H{\alpha}}$ line: when the system is in the soft-intermediate state \textendash \ with high X-ray flux \textendash \ the emission from the $\mathrm{H{\alpha}}$ line and the wind mass loss rate are lower. Follow-up studies found an orbital variability in optical \citep{2008AJ....136..631Y}, UV \citep{2008ApJ...678.1237G, 2008ApJ...678.1248V}, and X-ray spectra \citep{2015A&A...576A.117G, 2023A&A...680A..72H}. At the inferior conjunction of the star (when the star is in front of the black hole in the line of sight), the lines are more absorbed than at the superior conjunction. \cite{2023A&A...680A..72H} explain these deeper absorption features as being due to the presence of clumps in the outer region of the wind passing through the line of sight, which is commonly observed in line-driven winds \citep{1988ApJ...335..914O, 1997A&A...322..878F, 2006A&A...454..625P, 2012A&A...541A, 2013A&A...559A, 2013MNRAS.428.1837S, 2018A&A...611A..17S}. 
    Understanding the wind variability over changes in the X-ray spectral states is crucial to quantifying the variability in the accreted stream of matter, which probably also impacts the black hole's accretion rate.\\
    
    In this paper we present simultaneous X-ray and optical observations of the HMXB Cyg X-1 from 2022-2023, in both the low-hard and high-soft-intermediate X-ray spectral states, using high-resolution spectroscopy. We analyzed the Balmer line profiles' dependance on the orbital phase and the X-ray spectral state of the system. 
    In addition to the prominent $\mathrm{H{\alpha}}$ line that the previous studies focused on, we analyzed the He II line observed in the accreted stream of matter located between the star and the accretion disk -- also known as the "focused wind" \citep[][and others]{1982ApJFriend...261..293F, 1986ApJ...304..371G,1986ApJ...304..389G, 1998ApJ...506..424S, 2005ApJ...620..398M, 2009ApJ...690..330H}. 
    The analysis of the line profile variations provides the radial velocities and the line intensities, which are necessary for determining the physical processes governing the variability in the wind's emission. To understand the impact of X-ray spectral state transitions on the accreted wind in HMXBs, we separated the emission originating from the star's photospheric region and the focused wind. \\
    \indent The content of this paper is as follows: In Sect. \ref{observations} we present our monitoring campaign in the optical, the simultaneous X-ray observations, and the data reduction.
    In Sect. \ref{disentangling method} we describe the method of Fourier disentangling \citep{1995A&AS_petr..114..393H, 1998vsr..conf..111H, 2006Ap&SS.304..337H, hadrava_disentangling_2009}, which we used to isolate the focused wind's component and study its variability over a state transition as well as to study the line's parameter variations. 
    Section \ref{results} shows the results from the disentangling, such as the radial velocity calculations and the X-ray/optical anticorrelation. We discuss the line profile variation from the photosphere and the focused wind, together with the consequences of the X-ray/optical anticorrelation, in Sect. \ref{Discussion}. Our conclusions are presented in Sect. \ref{discussion and ccl}. 


\section{Observations and data reduction}\label{observations}
\subsection{Optical observations with the Perek telescope}

The Perek 2m telescope \textendash \ located at the Ond\v{r}ejov Observatory in the Czech Republic \textendash \ monitored Cyg X-1 in the low-hard state for 5 months in 2022 and during its transition to the high-soft-intermediate state for 4 months in 2023. We used two fiber-fed medium and high-resolution spectrographs mounted on the Perek 2m telescope. We performed high-resolution spectroscopy with the Ond\v{r}ejov Echelle Spectrograph \citep[OES;][]{2004_OES, Kabath_2020} with a resolving power of 51600 around $\mathrm{H\alpha}$ (6562 $\mathrm{\AA}$) and a spectral coverage between 3753 and 9195 $\mathrm{\AA}$. The OES was used for a total of about 27 hours (97.089 ks) in 2022 and 11.2 hours (40.200 ks) in 2023. A total of 38 spectra were acquired -- 27 in the low-hard state and 11 in the high-soft-intermediate state. \\
\indent In complement  to the OES, we used the medium-resolution single-order spectrograph \citep[CCD700,][]{2002PAICz..90...22S}. The spectra were taken in the red spectral region around $\mathrm{H\alpha}$ (6562 $\mathrm{\AA}$), providing a resolution of about 13000. The lower-resolution spectra of the CCD700, compared to that of the OES, are principally used for calibration and normalization. A total of about 17 hours (60.862 ks) were acquired in 2022 and 11.4 hours (41.070 ks) in 2023. A total of 29 spectra were obtained with the CCD700 -- 17 in the low-hard state and 12 in the high-soft-intermediate state.  The observation logs of both spectrographs in 2022 and 2023 are summarized in Tables~\ref{ondrejov2022} and \ref{ondrejov2023}.
We reduced the OES spectra using the IRAF\footnote{IRAF is distributed by the National Optical Astronomy Observatories, operated by the Association of Universities for Research in Astronomy, Inc. under a cooperative agreement with the National Science Foundation.} \citep[Image Reduction and Analysis Facility;][]{Tody_1986SPIE..627..733T, Tody_1993ASPC...52..173T} package and a dedicated semi-automatic pipeline \citep{cabezas_2023_10024183}\footnote{\url{https://zenodo.org/records/10024183}}, which incorporates a full range of standard procedures for the Echelle spectra reduction: bias correction, flat-fielding, wavelength calibration, heliocentric velocity correction, and continuum normalization. We reduced the CCD700 spectra using the following standard IRAF tasks: bias subtraction, creation of 1D spectra, flat-fielding, wavelength calibration and heliocentric velocity correction.

\subsection{Optical photometry from Las Cumbres Observatory}

Optical monitoring of Cyg~X-1 was performed with the Las Cumbres Observatory \citep[LCO;][]{brown_2013_lco} 1m robotic telescopes located at the McDonald Observatory (Texas, USA) and at the Teide Observatory (Tenerife, Spain), from 2023 June 02 (MJD 60097.34) to 2023 September 28 (MJD 60215.84), using $B$, $V$, $r^{\prime}$ and $i^{\prime}$ filters.
All the observations were done with a 2 second exposure time to avoid saturation due to the brightness of the system. Then, the images were processed and analyzed by the X-ray Binary New Early Warning System (XB-NEWS) pipeline \citep[see, e.g.,][]{russell_2019_precursors, goodwin_2020_saxj1808}, carrying out the following tasks:

\begin{enumerate}
\item Download and calibrate images from the LCO database (i.e., bias, dark, and flat-field images).
\item Perform data reduction and reject any images of poor quality.
\item Perform astrometry using \textit{Gaia} DR2\footnote{\url{https://www.cosmos.esa.int/web/gaia/dr2}} positions.
\item Carry out multi-aperture photometry \citep{stetson_1990_map}.
\item Solve for photometric zero-point offsets between epochs \citep{bramich_2012_photometry}.
\item Flux-calibrate the photometry using the ATLAS All-Sky Stellar Reference catalogs (ATLAS-REFCAT2) \citep{tonry_2018_atlas}\footnote{\url{https://archive.stsci.edu/prepds/atlas-refcat2/}}, which includes the AAVSO Photometric All-Sky Survey (APASS), the Panoramic Survey Telescope and Rapid Response System (Pan-STARRS DR1), and other catalogs.
\end{enumerate}
We rejected any magnitude with an uncertainty above 0.25 mag. After XB-NEWS data processing, we had 17, 15, 14, and 15 data points in, respectively, $B$, $V$, $r^{\prime}$, and $i^{\prime}$.

\subsection{Optical photometry from TESS}

To check the optical variability of Cyg~X-1, we analyzed data from the Transiting Exoplanet Survey Satellite \citep[TESS;][]{Ricker2015} between July (MJD 59784.5) and August 2022 (MJD 59809.5). We processed the data with the TESS Science Processing Operations Center \citep[SPOC;][]{Jenkins2016} and the quick-look pipeline \citep[QLP;][]{Huang2020a,Huang2020b}. We used the \textsc{Lightkurve v2.4} software to download the data \citep{Lightkurve2018,Barentsen2020}. We checked long-cadence SPOC and QLP data (1800+600\,seconds, sectors 14+54+55) and short-cadence SPOC data (120\,seconds, sectors 14+54+55+74). The two datasets are of similar quality and give consistent results. 

\begin{figure}[h]
   \centering
   \includegraphics[width=0.4\textwidth]{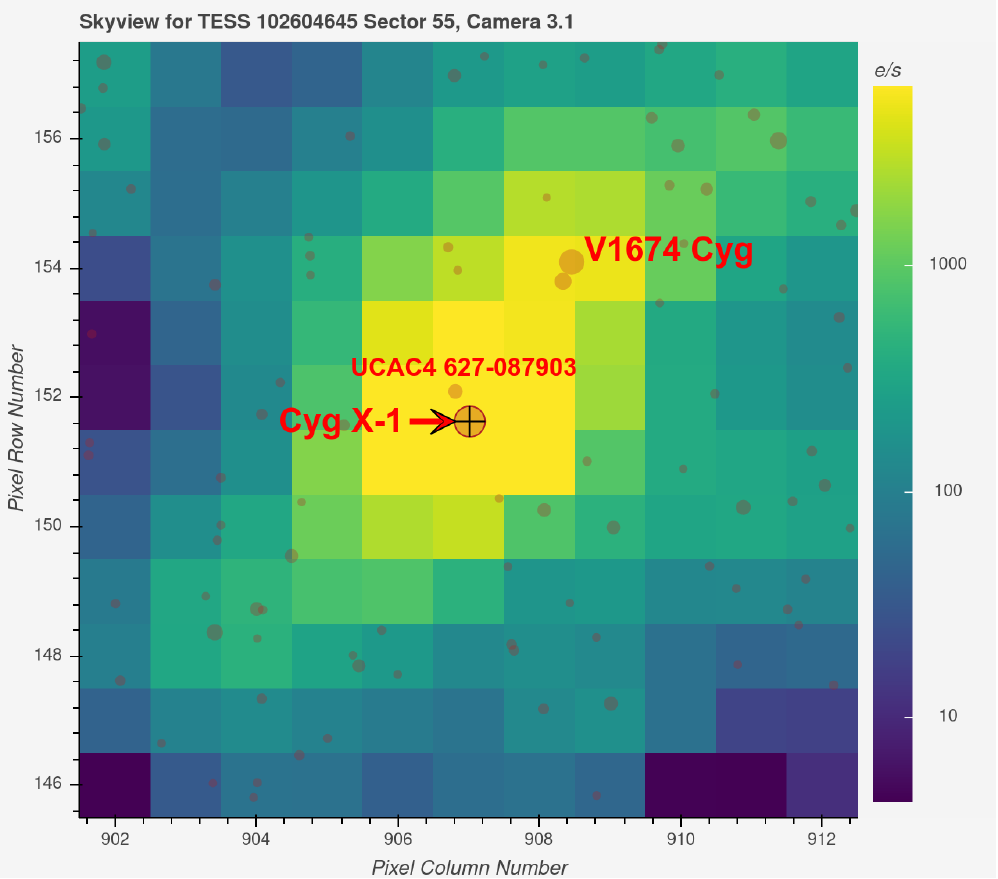}
   \caption{Vicinity of Cyg~X-1 on the TESS target pixel color-coded by the flux in pixels and showing nearby stars from \textit{Gaia} DR3 \citep{GaiaDR3}. The most likely contaminants are labeled. The faintest stars are of $V\approx 18$\,mag.}
              \label{Fig:Chart}
\end{figure}
\noindent The angular resolution of the TESS satellite is only 21"/px \citep{Ricker2015}. Therefore, we can expect a strong contamination by nearby sources in the crowded field of Cyg~X-1. There are several dozen nearby stars (see Fig.~\ref{Fig:Chart}), with the brightest being the spectroscopic binary V1674\,Cyg (54" from Cyg~X-1, $V=10.07$\,mag) and the closest being UCAC4\,627-087903 (9", $J=10.285$\,mag). However, there is no doubt that the observed variations originate from Cyg~X-1. First, the custom-aperture photometry of V1674\,Cyg and UCAC4\,627-087903 do show the same variation as Cyg~X-1 but with much lower amplitude. Second, the observed variations have a period equal to that of Cyg~X-1, previously determined by spectroscopy (see the frequency spectrum in the top panel of Fig.~\ref{Fig:TESS}).

\subsection{X-ray observations with MAXI}

The Monitor of All-sky X-ray Image \citep[MAXI;][]{2009PASJ...61..999M} on board the International Space Station (ISS) covers the entire sky in the 92 minutes it takes for the ISS to orbit the Earth. Two X-ray detectors are used on MAXI: a gas proportional counters -- the Gas Slit Camera \citep[GSC; 2--30 keV;][]{2002SPIE.4497..173M, 2011PASJ...63S.635S, 2011GSC_PASJ...63S.623M} -- and an X-ray CCD -- the Solid-state Slit Camera \citep[SSC; 0.5--12 keV;][]{2005NIMPA.541..350K, 2010PASJ...62.1371T, 2011PASJ...63..397T}. MAXI continuously monitors Cyg X-1 over the 2-20 keV energy range. Therefore, we have simultaneous X-ray observations with the optical monitoring of the source. We obtained the light curves and X-ray spectra using the "on-demand process" of MAXI official web page\footnote{\url{http://maxi.riken.jp/mxondem/}}, with a light curve time bin of 0.1 days ($\mathrm{\approx 2 \ hours}$).

\section{The method of Fourier disentangling}\label{disentangling method}

\begin{figure}[h]
    \centering
    \includegraphics[width=9cm]{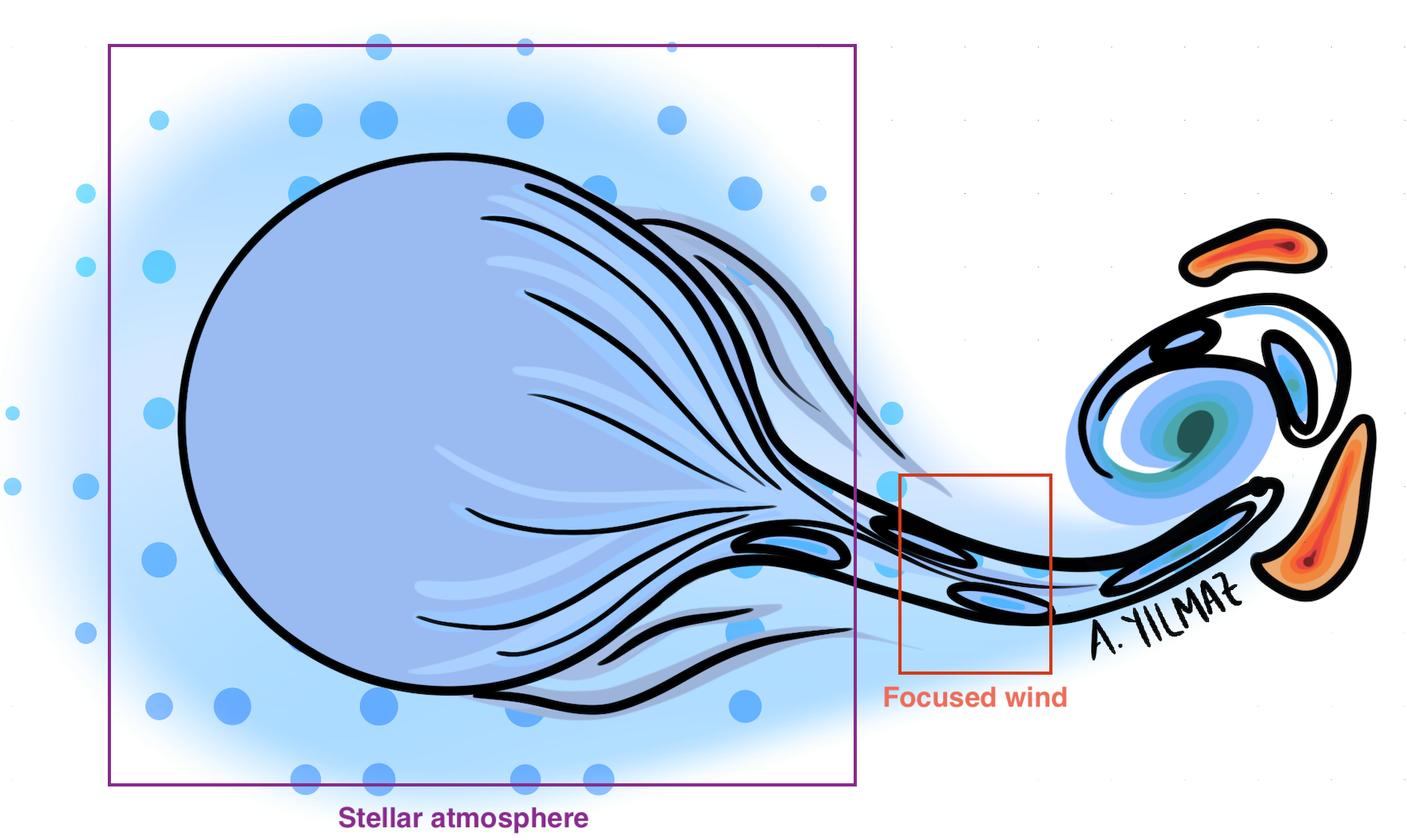}
    \caption{Artist's impression of Cyg X-1. The purple and orange boxes show the two components isolated in the disentangling method. We note that this figure is only illustrative and does not represent a physical solution from the Fourier disentangling method. }
    \label{fig:sketch anastasiya}
\end{figure}
\noindent We used the method of Fourier disentangling to separate the stellar atmosphere and the focused wind spectral signatures from the composite optical spectra. 
Developed by Petr Hadrava in 1995 \citep{1995A&AS_petr..114..393H,2006Ap&SS.304..337H, 2016ASSL..439..113H}, 
the method of Fourier disentangling only uses the orbital motion of the different components in the system to decompose the spectrum of a multiple system into sub-spectra for each component of the system. It determines the orbital parameters of the global system and gives the intensity of the lines (in emission and in absorption) for each separated component. The method combines multiple Doppler shifted spectra at different orbital phases and applies the least-square fit of their Fourier transform. The Doppler shift of the spectra enables the determination of the radial velocities of each component, constraining the orbital motion of the system. \cite{1994A&A...281..286S} developed the first version of the disentangling method using the singular value decomposition technique. They compared the line profiles of the system at different orbital phases (out of eclipse) and solved a linear set of equations in the wavelength domain (in opposition to the Fourier domain, used in the version of the method of Fourier disentangling).

In Cyg X-1, most of the optical emission comes from the star: \cite{1978SvAL....4..292B} estimated the accretion disk to only contribute to 2\% of the total optical emission. Figure \ref{fig:sketch anastasiya} shows an artist's representation of Cyg X-1.  
The black hole is surrounded by the accretion disk -- mainly emitting in X-rays -- and the supergiant's wind -- emitting in UV and optical -- is ionized by the X-ray emission. The stellar wind faces strong inhomogeneities because of the X-ray irradiation, appearing as clumps in the outer atmosphere and in the focused wind region. Regarding the accretion disk emission, no features were found in the optical spectra, which can be explained by the lines being highly broadened because of the disk's rotation. \\

For the Fourier disentangling, we used the code \texttt{KOREL} \citep{2004PAICz..92...15H} to determine in which part of the circumstellar matter the optical emission originates and separate it from the spectrum of the donor star. 
A successful disentangling requires more spectra than the number of components to be disentangled. First, we considered the spectrum of a system with $n$ components observed at the time $t$:
\begin{equation}\label{eq1}    
I(x,t) = \sum^{n}_{j=1} \ I_{j}(x) \ \ast \ \Delta_j(x,t,p) \; ,
\end{equation} 
which is a superposition of the different components' intrinsic spectra ${I_{j}(x)|^n_{j=1}}$. Each component's spectrum ${I_{j}}$ is supposed constant, and the spectral line's variability is usually considered as a second-order perturbation that can be neglected.
Each spectrum is convolved in the logarithmic wavelength scale $x=c \ \times \ \ln(\lambda)$ with a broadening function $\Delta_j(x,t,p)$, introducing variability to the spectrum.

In the code \texttt{KOREL}, we have as input: the orbital period (fixed at 5.599829 days), the eccentricity (fixed at 0), the periastron longitude (fixed at $-90$°) and the respective time variations of the above parameters set to 0. In this case, the only time variation of the component’s spectra is due to the Doppler shift. Additional inputs are the total number of spectra used alongside the wavelength of interest.
To obtain the values of the parameters $p$, we then performed a least-square fit of the Fourier transform of all observed spectra together. Simultaneously, we least-square-fit the Fourier modes $\Tilde{I}_j$ of the intrinsic spectra, which are simple multiplicators of $\Tilde{\Delta}_j$. For each mode of the spectra, we get a separate set of linear equations of order $n\times n$.

In this study we assumed that the only time variability of the components is given by the Doppler shift through the radial velocities $v_j(t,p)$ and the strength factors $s_j(t)$ (also called the s-factors):
\begin{equation}\label{strength_factor}    
\Delta_j (x, t, p) = s_{j}(t) \times \delta (x - v_j(t,p))\; , 
\end{equation}
which is \begin{equation}\label{fourier_strength_factor}   \Tilde{\Delta}_j (y, t, p) = s_{j}(t) \times \e^{iy v_j(t,p)} \end{equation} in the Fourier space. The radial velocities $v_j$ in the shifted Dirac $\delta$-function are computed according to the Keplerian orbital motion with the parameters $p$ \citep[see][]{hadrava_disentangling_2009}. The amplitudes and times of periastron passage of the separated components are converged during the calculation. The line-strength factors $s_j$ were originally introduced into the disentangling method to improve spectral fitting during eclipses in eclipsing binaries. They give the ratio between the line strength in a given exposure $I(x,t_j)$ over the mean spectrum $\langle I(x)\rangle$:
\begin{equation}\label{def s-factors}    
s_j = \frac{I(x, t_j)}{< I(x) >} \; ,
\end{equation} 
with $\mathrm{I(x, t_j) >0}$ for emission lines and $\mathrm{I(x, t_j) <0}$ for absorption lines.

In the case of Cyg X-1, we separated the stellar atmosphere emission from the focused wind emission (see Fig. \ref{fig:sketch anastasiya}), determined the amplitudes and phases of their radial velocities, and quantified the changes of the emission-lines strengths with the X-ray state of the system. The location of the focused wind within the system is solved by calculation of the phase shift of its radial velocities' amplitudes with respect to the stellar atmosphere. In a more general case, the method of disentangling can lack precision if the orbital parameters are not correctly constrained, if the phase coverage is too sparse, and if the spectra are too variable or too noisy. For Cyg X-1, the disentangling is quite accurate considering that the source is particularly bright and the orbital parameters well constrained. Furthermore, the high-resolution spectrograph covers a wide range of orbital phases in both the hard and soft X-ray spectral states.

\section{Results}\label{results}
\subsection{Timing analysis}

    \begin{figure}[h]
      \centering
      \includegraphics[width=9.2cm]{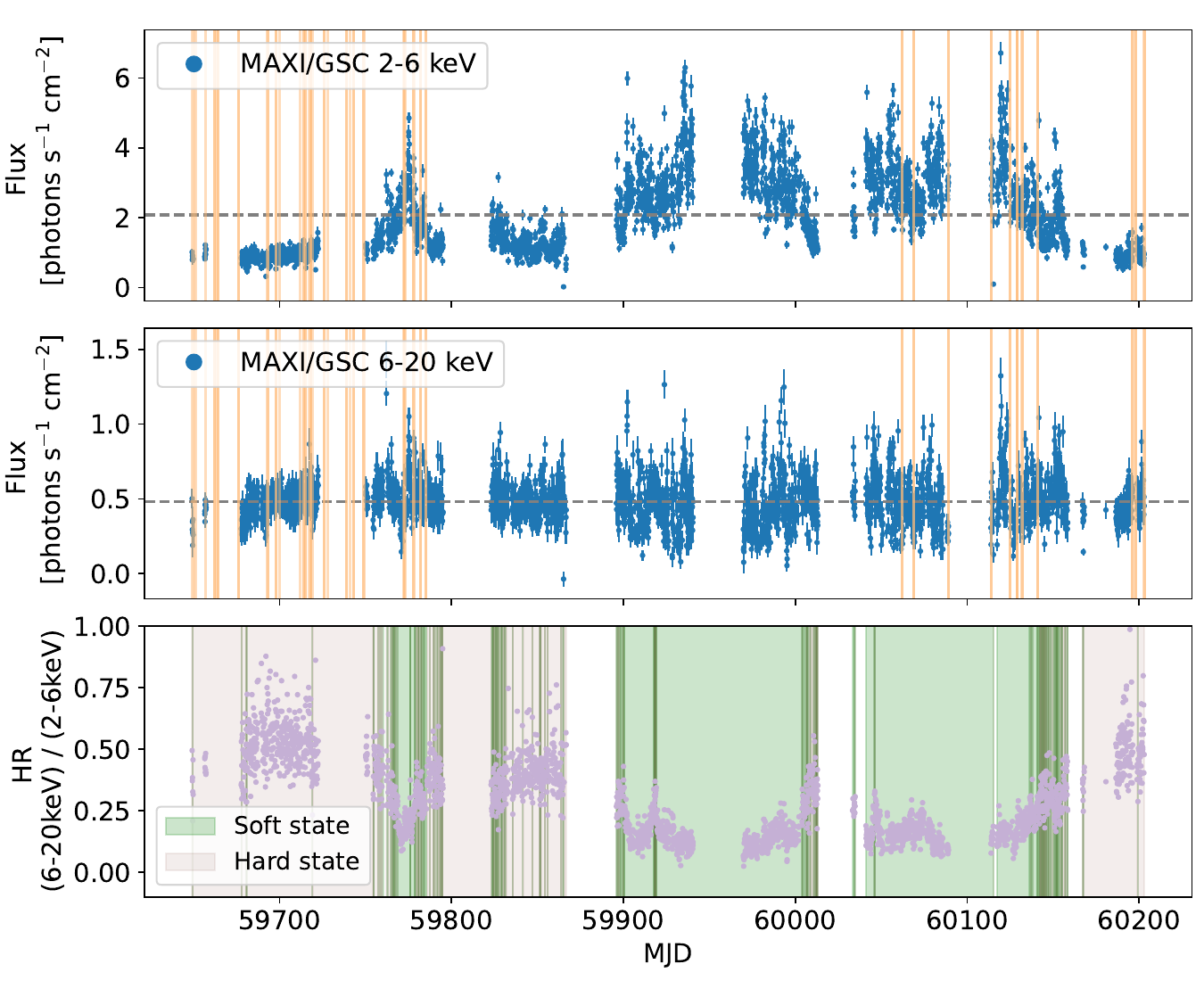}
      \caption{MAXI/GSC X-ray light curves and hardness ratio calculated between March 2022 and September 2023 over the energy range 2--20 keV, with a 0.1 day time bin ($\mathrm{\approx 2 \ hours}$). \textit{Upper panel}: Light curve in the 2--6 keV energy range. \textit{Middle panel}: Light curve in the 6--20 keV energy range. In the upper and middle panels, the orange vertical lines indicate the simultaneous optical observations with the Perek telescope, and the horizontal dashed gray lines indicate the mean value of the X-ray flux. \textit{Lower panel}: Hardness ratio calculated as the ratio between the count rates in the 6--20 keV energy range over the count rates in the 2--6 keV energy range. The green-shaded regions show the period when the system is considered to be in the soft-intermediate X-ray state and the brown-shaded regions show when the system was in the hard state. 
              }
         \label{fig:MAXI lc}
   \end{figure}
MAXI (2--20 keV) monitored Cyg X-1 simultaneously with the optical observations. Figure \ref{fig:MAXI lc} shows the X-ray light curves from MAXI/GSC in two energy bands and the hardness ratio defined as the ratio between count rates in the 6--20 keV energy range over count rates in the 2--6 keV energy range. From top to bottom are the light curves in the 2--6 keV energy band, the 6--20 keV energy band, and the hardness ratio. The simultaneous optical observations from the Perek telescope are indicated in the figure as the orange vertical lines, and the shaded regions in the hardness ratio plot show the soft (shaded green) and hard (shaded brown) states. The spectral states are identified by variations in the X-ray flux light curves. The upper panel of Fig. \ref{fig:MAXI lc} represents the soft energy band (below 6 keV) and the middle panel represents the hard energy band (above 6 keV). The X-ray flux is more variable in the 2--6 keV energy range than the 6--20 keV energy range: in the soft band, the maximum flux gradient is 6 $\mathrm{photons \ s^{-1} \ cm^{-2}}$ versus 1.2 $\mathrm{photons \ s^{-1} \ cm^{-2}}$ in the hard band. Hence, the system is the most variable in flux at low energy, below 6 keV.

In the soft band, between the beginning of the observations and MJD 59723 the flux is low and varies between 0.6 and 1 $\mathrm{photons \ s^{-1} \ cm^{-2}}$. In the hard band, the flux varies between 0.2 and 0.7 $\mathrm{photons \ s^{-1} \ cm^{-2}}$. Over this time, the hardness ratio remains relatively constant at 0.5, indicating that the emission is primarily dominated by the Comptonized radiation from the corona, characteristic of the low-hard state. Cyg X-1 spent 90 \% of its time in this state between 1996 and 2000 \citep{2006A&A...447..245W} but started to reverse the trend in the early 2000 when it only spent approximately $62 \%$ of its time. During the time of our observations, the system spent 60\% of its time in the high-soft-intermediate state and only 40\% in the low-hard state.\\
\indent At MJD 59771 the hardness ratio drops from 0.5 to 0.1 -- the emission becomes softer because of the soft electrons coming from the accretion disk. The drop in the hardness ratio translates into a peak in the 2--6 keV energy band, increasing the flux from 1.0 to 5.1 $\mathrm{photons \ s^{-1} \ cm^{-2}}$. The system transitioned from a low-hard state to a high-soft-intermediate state, with the X-ray emission dominated by the thermal emission from the accretion disk. The high flux is maintained for a couple of days and the system goes back to a low-hard state at MJD 59795 with an average flux of 1 $\mathrm{photons \ s^{-1} \ cm^{-2}}$ in the 2--6 keV energy range. \\
\indent At MJD 59823 the hardness ratio stays relatively constant at 0.4 for 40 days, along with the flux over the entire energy range. Two months later (MJD 59895), the hardness ratio drops to 0.1. The dip in the hardness ratio marks the beginning of a period where the source's emission becomes softer. In the soft band, the X-ray flux varies between 1.5 and 6 $\mathrm{photons \ s^{-1} \ cm^{-2}}$, but remains constant around 0.5 $\mathrm{photons \ s^{-1} \ cm^{-2}}$ in the hard band (within the errors). Overall, the hardness ratio remains below 0.4 except at MJD 60010--60012 where the emission becomes harder for 2 days. The lack of observations between MJD 60014 and 60032 prevents the exact tracking of the X-ray flux variation. Nevertheless, the system sustains a high flux on average (above 1.8 $\mathrm{photons \ s^{-1} \ cm^{-2}}$) until MJD 60137 where it decreases below this limit, flares at MJD 60150, and drops again. At MJD 60155, the X-ray emission decreases, the hardness ratio increases (from 0.3 to a maximum value of 0.9) and the system stabilizes in a low-hard state. In the following parts of the paper, we consider the system to be in a high-soft-intermediate state when the average flux is above 1.5~$\mathrm{photons~s^{-1}~cm^{-2}}$ and the hardness ratio is below 0.3.\\

\begin{figure}[h]
    \centering
    \includegraphics[width=7cm]{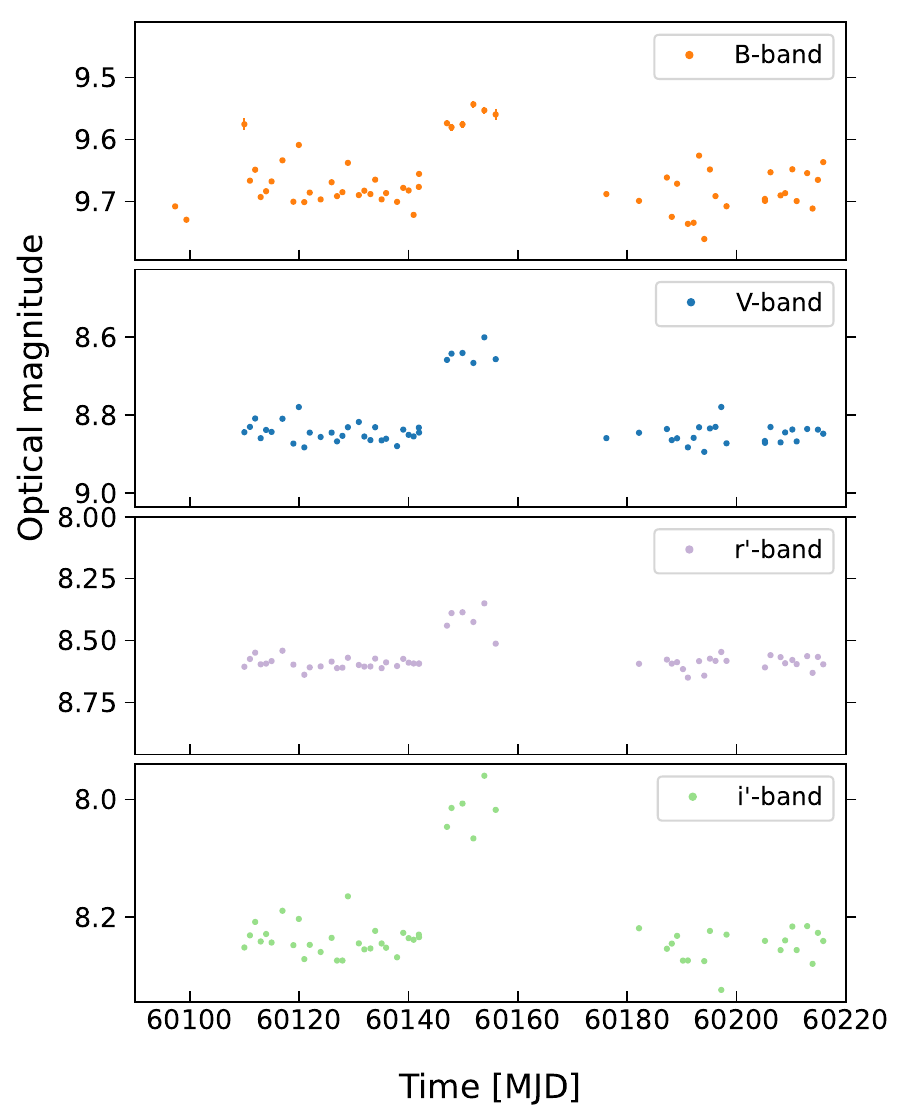}
    \caption{Photometric light curves in the B, V, $\mathrm{r '}$, and $\mathrm{i'}$ bands from LCO taken in 2023 in the high-soft-intermediate state.}
    \label{fig:LCO lc}
\end{figure}
%

In the optical, the observations were performed in the three principal phases of the system: in the lowest X-ray flux, during the flaring event at MJD 59775 and at high X-ray flux. Complementary to the X-ray light curves, Fig. \ref{fig:LCO lc} shows the photometric light curves from the LCO monitoring of the source in 2023 in the high-soft-intermediate state, in 4 different bands: the B, V, $\mathrm{r '}$ and $\mathrm{i'}$ bands . The source is stable on average in the four bands with 0.1 mag fluctuations between MJD 60110--60142 and MJD 60176--60215. The first interval MJD 60110--60142 corresponds to the transition from the high-soft-intermediate state to the low-hard state in X-rays and the second interval MJD 60176--60215 corresponds to the low-hard state (see Fig. \ref{fig:MAXI lc}). Thus, the high and the low X-ray flux can simultaneously occur with a constant optical magnitude. Furthermore, we observe an increase in magnitude in the B, V, $\mathrm{r '}$, and $\mathrm{i'}$ bands at MJD 60142 of, respectively, 
0.15, 0.18, 0.25, and 0.25, lasting minimum 10 days. This jump corresponds to the transition from the high-soft-intermediate state to the low-hard state. Nevertheless, the low-hard state is observed during the constant optical magnitude and during the jump. 
\begin{figure}[h]
   \centering
   \includegraphics[width=0.4\textwidth]{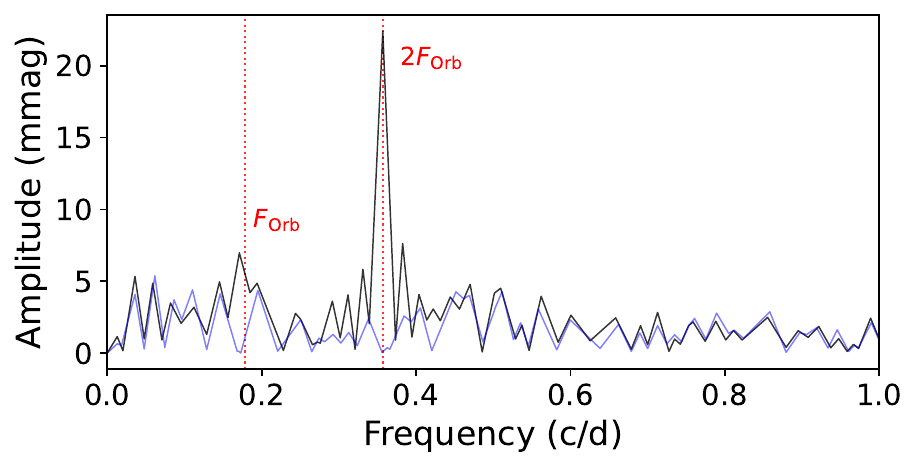}
   
   \includegraphics[width=0.4\textwidth]{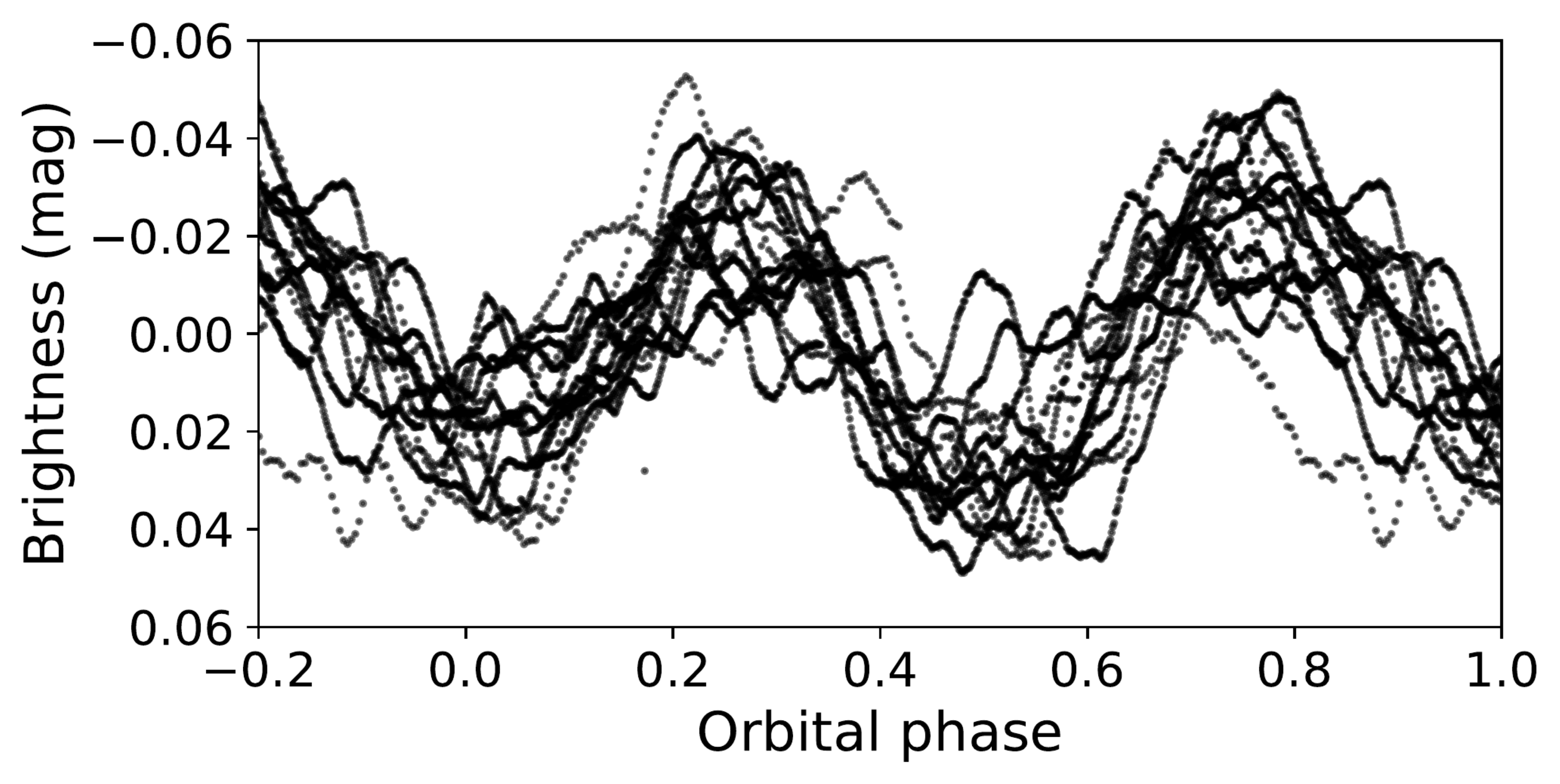}
   
   \includegraphics[width=0.4\textwidth]{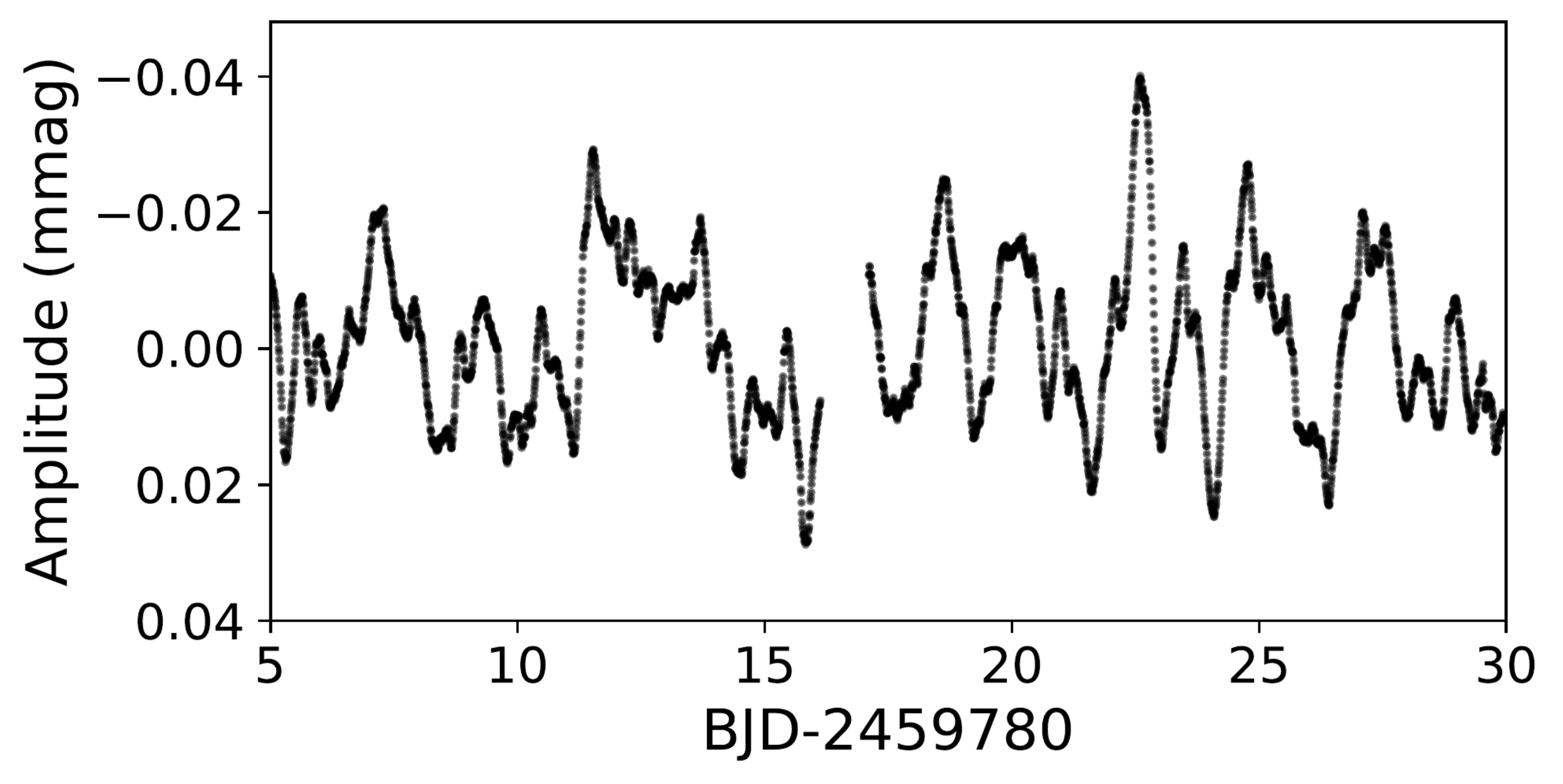}
   \caption{{\it Top panel:} Fourier transform of the full QLP dataset showing the orbital frequency and its harmonic (in black) and the residuals after pre-whitening (in blue). {\it Middle panel:} Data phase-folded with the orbital period and ephemeris from \cite{2003ApJ...583..424G}. {\it Bottom panel:} Stochastic variations after the ellipsoidal variation were removed.}
              \label{Fig:TESS}
\end{figure}

\noindent Complementary to the LCO photometric analysis, Fig. \ref{Fig:TESS} shows the photometric measurements from TESS starting four days after the flaring event, at MJD 59779.5. While the middle panel of Fig.~\ref{Fig:TESS} shows the ellipsoidal variations caused by the tidal deformation of the primary star, \citep[already studied by][]{Avni1975,Sorabella2022}, the light curve in the bottom panel shows stochastic or semi-regular variations without measurable periodicity, which is typical for O-type stars. It is important to note that we did not identify any correlation between the TESS light curve and the soft and hard X-ray spectral states, which is consistent with the results from the LCO photometry.

\subsection{Disentangling of the hydrogen lines}

\begin{figure*}[t]
    \centering
    \includegraphics[width=6cm]{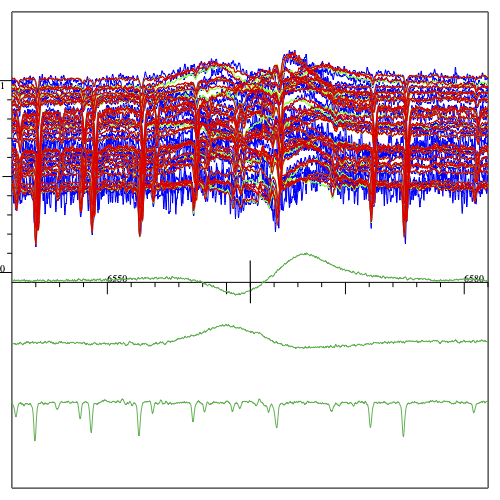}
    \includegraphics[width=6cm]{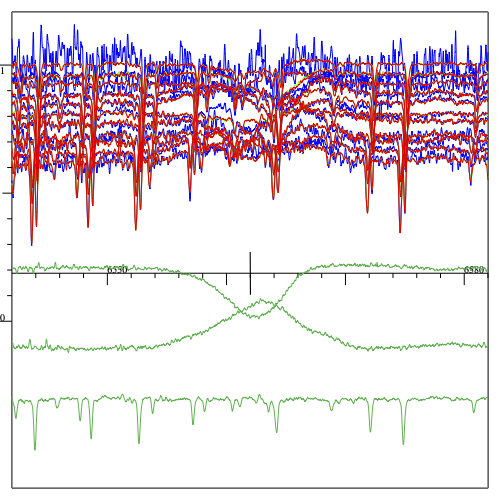}
    \caption{Disentangled spectra of $\mathrm{H{\alpha}}$ with a central wavelength at $\mathrm{6562 \ \AA}$, indicated with the vertical black line. \textit{Left}: Disentangled spectra from the 2022 epoch. \textit{Right}: Disentangled spectra from the 2023 epoch. The top of each plot combines the OES observed spectra (blue) and the synthetic spectra from the \texttt{KOREL} radial velocity solutions (red) at different orbital phases. The bottom three green lines show the disentangled lines: the stellar atmosphere, the focused wind, and the Earth's atmosphere (from top to bottom).}
    \label{fig: disentangled Ha}
\end{figure*}

\begin{figure*}[t]
    \centering
    \includegraphics[width=6cm]{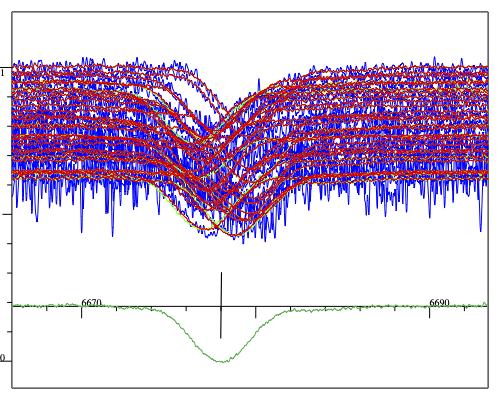}
    \includegraphics[width=6cm]{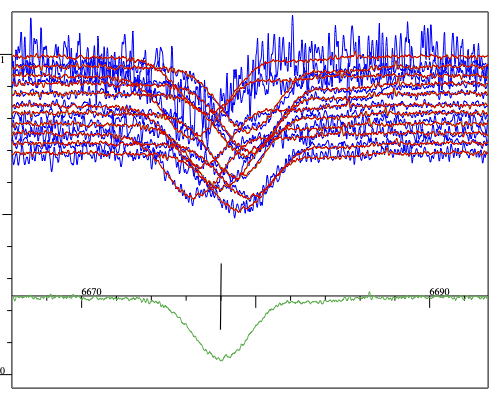}
    \caption{Same as Fig. \ref{fig: disentangled Ha} but with the disentangled spectra of He I with a central wavelength at 6678 $\mathrm{\AA}$, indicated with the vertical black line.
    We do not show the telluric lines from the Earth’s atmosphere as they are not considered in the analysis. We do not detect the focused wind spectral component. }
    \label{fig: disentangled HeI6678}
\end{figure*}

Over the wide spectral range the OES covers (3753--9195 $\mathrm{\AA}$) we observed the Balmer lines of hydrogen and helium and heavier metals such as nitrogen and silicon. We then combined the OES spectra taken at different orbital phases and used the method of Fourier disentangling to separate the spectral lines coming from the star's atmosphere, the focused wind and the telluric lines from the Earth's atmosphere. We then determined the orbital parameters using the \texttt{KOREL} code. Figure \ref{fig: disentangled Ha} shows a typical output of the \texttt{KOREL} code. This figure represents the OES disentangled spectra for the $\mathrm{H{\alpha}}$ line, at the central wavelength $\mathrm{6562 \ \AA}$, in the low-hard (left) and the high-soft-intermediate (right) X-ray spectral states (in the appendix we show the Fourier disentangling of $\mathrm{H{\beta} \ \lambda 4860}$ and $\mathrm{He \ I \ \lambda 5875)}$. The blue lines are the normalized observed spectra. The red lines represent the fit by the superposition of the separated profiles with the radial velocities at the time of exposure computed according to the orbital parameters -- they are the synthetic spectra from the Fourier disentangling solutions. The green lines show the resulting separated spectra for (from top to bottom): the star's atmosphere, the focused wind and the Earth's atmosphere. During the low-hard state, we combined 27 spectra versus 11 spectra in the high-soft-intermediate state. 
The disentangling of the spectra in the low-hard state is more precise and less noisy than in the high-soft-intermediate state due to the larger number of observed spectra for this state. Furthermore, the focused wind in Cyg X-1 is not firmly localized in the velocity space, it has an anisotropic distribution and is variable in time. As a result, part of its emission (reaching up to about 3\% of the continuum level compared to about 14\% of the focused wind emission) leaked into the separated spectrum of the telluric lines, which also has a small radial velocity amplitude. Thus, we used a template to constrain the telluric spectra and remove the leaked emission. The residuals between the original and separated spectra are calculated by subtracting the synthetic spectra (red lines) from the observed spectra (blue lines) in Figs. \ref{fig: disentangled Ha}, \ref{fig: disentangled HeI6678}, and \ref{fig: disentangled HeII44686}. The residuals are on average on the order of $\mathrm{10^{-2}}$ in units of the continuum level.\\
\indent In the low-hard state (left panel in Fig. \ref{fig: disentangled Ha}) we can identify the standard P-Cygni profile with a blueshifted absorption and a redshifted emission in the stellar atmosphere, which is a sign of an outflowing material. The focused wind has a blueshifted emission and no absorption. 
This opposite motion between the stellar atmosphere and the focused wind, here traced by the H$\alpha$ line, can also be observed  in the radial velocity curve for the helium lines in Fig. \ref{fig: RV from HeI6678 and HeII4686}. Compared to the low-hard state, the stellar emission feature is weaker in the high-soft state. Moreover, we observe differences in the broadening of the stellar atmosphere lines between the two spectral states. For instance, in the disentangling of the $\mathrm{H{\alpha}}$ line for the focused wind component, the average width is 4 $\mathrm{\AA}$ in the low-hard state versus 6 $\mathrm{\AA}$ in the high-soft-intermediate state.

\subsection{Disentangling of the helium lines}

\subsubsection{The case of $\mathrm{He \ I \ \lambda 6678}$}
\begin{figure}[h]
    \centering
    \includegraphics[width=7cm]{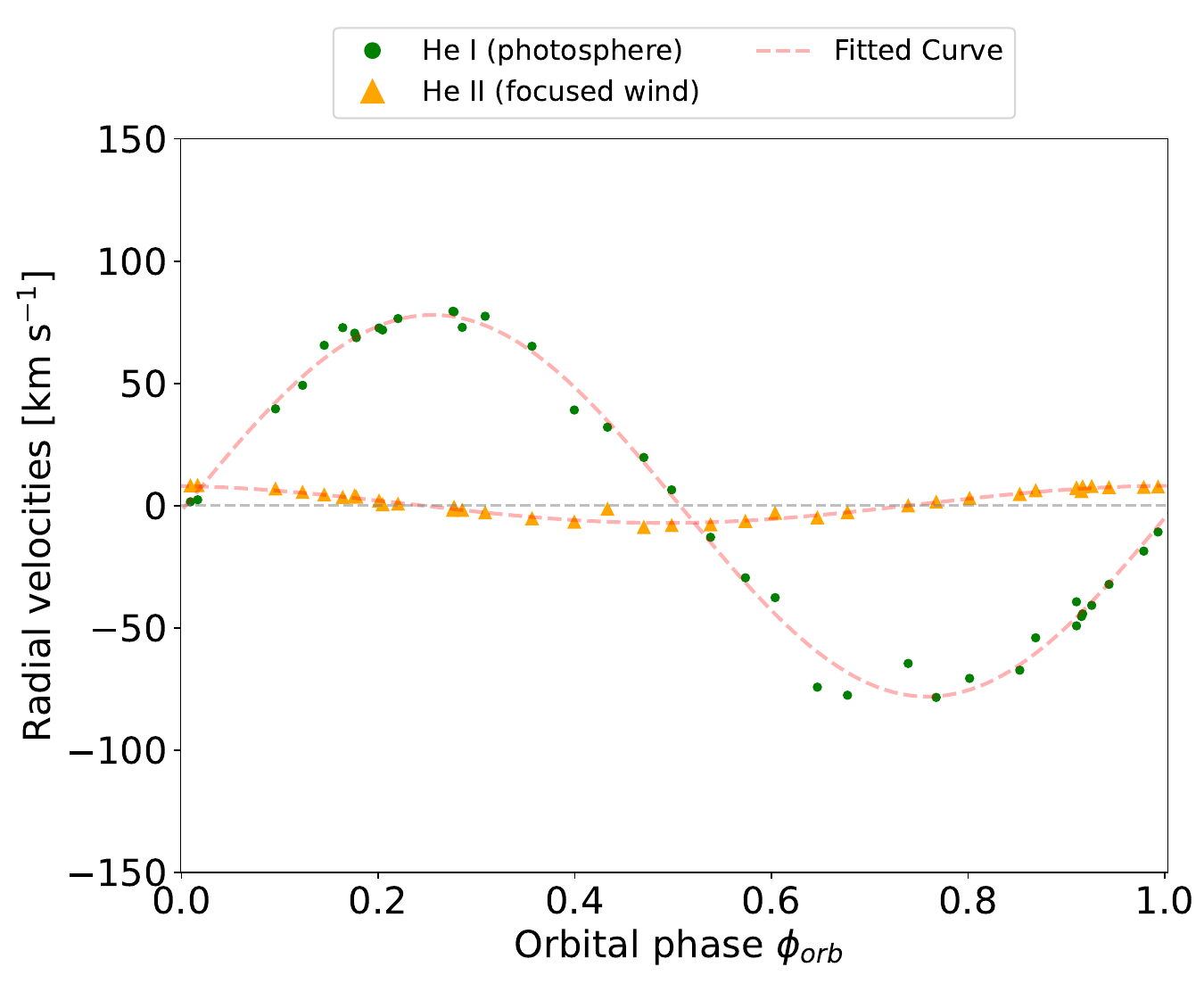}
    \caption{Radial velocity curve from the stellar atmosphere component of $\mathrm{He\ I \ \lambda 6678}$ (green circles) and the focused wind component of $\mathrm{He\ II \ \lambda 4686}$ (orange triangles) from the disentangling method. The orbital phases are calculated based on the ephemeris from \cite{2003ApJ...583..424G}: $\mathrm{2451730.449(8) + 5.599829(16) \times E.}$ The dashed red lines are the fitted functions.
    }
    \label{fig: RV from HeI6678 and HeII4686}
\end{figure}

Compared to $\mathrm{H{\alpha}}$, the helium line $\mathrm{He\ I \ \lambda 6678}$ only exhibits an absorption feature for the star's atmosphere in both states. We can see in Fig. \ref{fig: disentangled HeI6678} the disentangled $\mathrm{He \ I \ \lambda 6678}$ line in the low-hard (left) and high-soft-intermediate (right) state. Moreover, no significant $\mathrm{He\ I \ \lambda 6678}$ emission or absorption was detected in the focused wind for either state, making it an ideal candidate for reliably determining the star's radial velocities and orbital parameters. Figure \ref{fig: RV from HeI6678 and HeII4686} shows the radial velocity curve for the $\mathrm{He\ I}$ and $\mathrm{He\ II }$ lines. The data for the $\mathrm{He\ I}$ line (in green) are from the disentangled radial velocities calculation for the star's atmosphere component. We estimate the maximum radial velocity of the stellar atmosphere in absolute value at 79.1 $\mathrm{km \ s^{-1}}$. The $\mathrm{He\ II}$ radial velocities (in orange) are from the disentangled values for the focused wind. The dashed red lines represent the function\begin{equation}
    \mathrm{a_0 \times \sin{(2\pi({\bf \phi} \ + \psi))}}
        \label{eq: cosinus}
,\end{equation} 
with $\mathrm{a_0}$ the amplitude of the signal, $\mathbf{\phi}$ the orbital phase, and $\mathrm{\psi}$ the phase shift of the signal in units of the orbital phase. The fitted parameters for $\mathrm{He\ I}$ are $\mathrm{a_0 = 79.1 \ km \ s^{-1}}$ and $\mathrm{\psi = -0.01}$. For the $\mathrm{He\ II}$, $\mathrm{a_0 = 8.0 \ km \ s^{-1}}$, and $\mathrm{\psi = 0.26}$. The amplitude of the radial velocities of the He\,II line in the focused wind component is an order of magnitude lower than that of the He I line in the stellar atmosphere. Additionally, the two components are phase-shifted by approximately 0.25 (i.e., about a day and a half). As a result, the radial velocities of He\,II reach their maximum when the star is at the inferior conjunction, moving perpendicularly to the line of sight. At this point, the radial velocities of He I start to become positive.

\subsubsection{The case of $\mathrm{He \ II \ \lambda 4686}$}

\begin{figure*}[t]
    \centering
    \includegraphics[width=6cm]{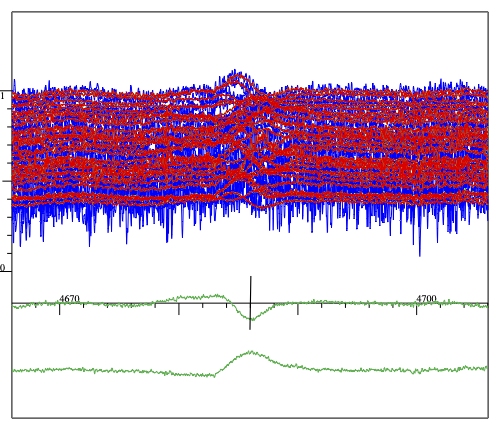}
    \includegraphics[width=6cm]{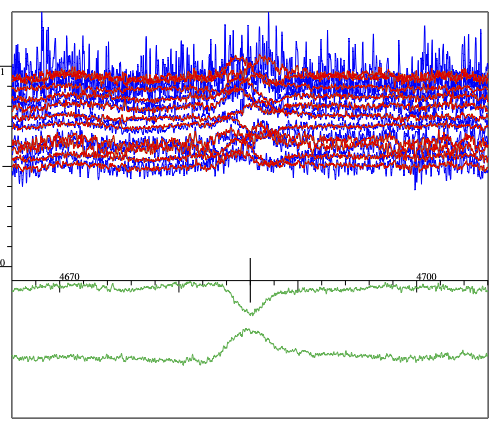}
    \caption{Same as Fig. \ref{fig: disentangled Ha} but with the disentangled spectra of He II with a central wavelength at 4686 $\mathrm{\AA}$, indicated with the vertical black line. We do not show the telluric lines from the Earth’s atmosphere as they are not considered in the analysis.}
    \label{fig: disentangled HeII44686}
\end{figure*}

\begin{figure}[h]
    \centering
    \includegraphics[width = 7cm]{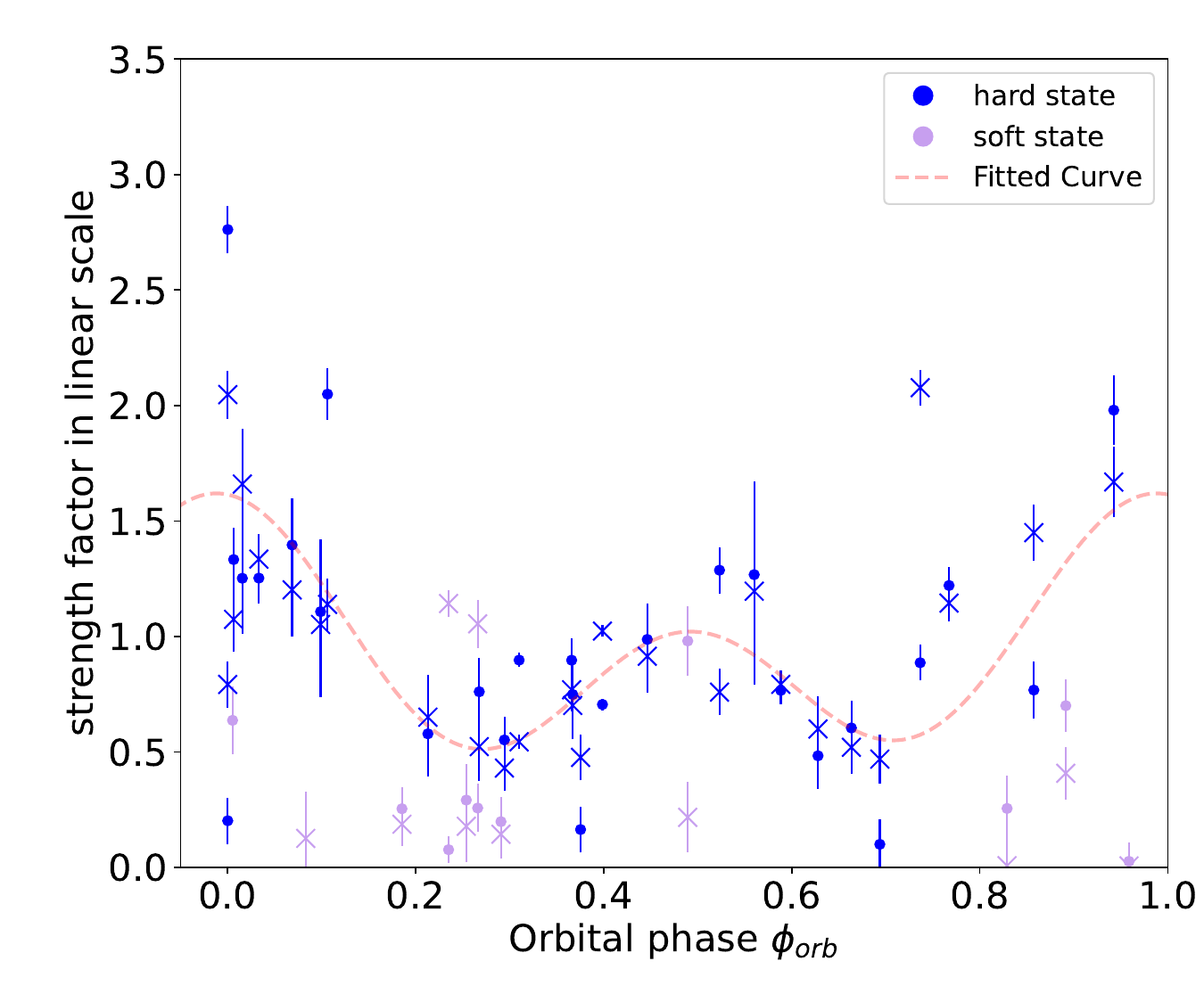}
    \includegraphics[width = 7cm]{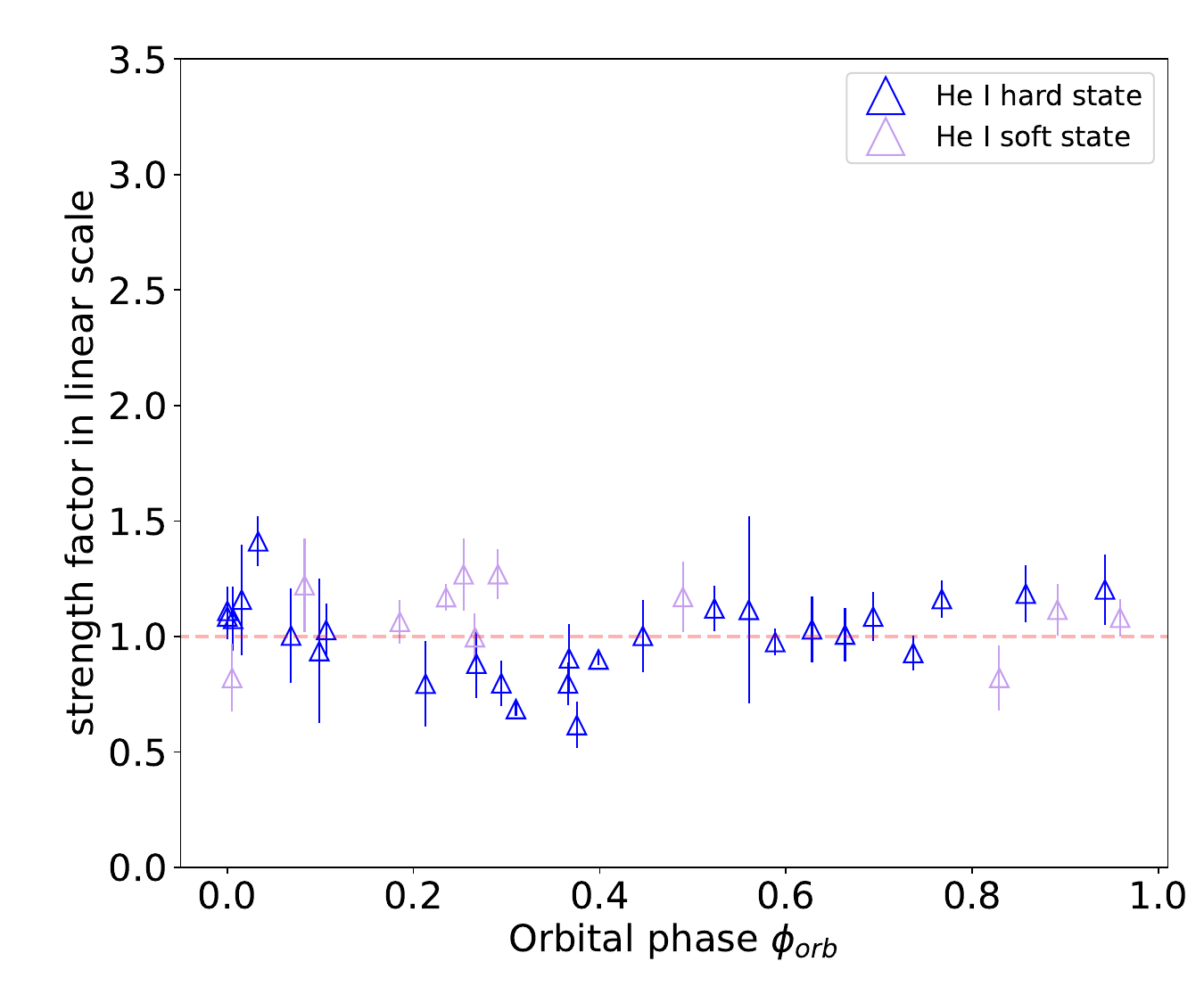}
    \caption{Evolution of the strength factors with respect to the orbital phase. \textit{Top}: For $\mathrm{H \alpha \ \lambda 6562}$ (circles) and $\mathrm{He \ II \ \lambda 4686}$ (crosses), for the stellar atmosphere component. The dashed red curve shows the best fit from the Fourier analysis. \textit{Bottom}: For $\mathrm{He \ I \ \lambda 6678}$, for the stellar atmosphere component. The horizontal dashed red curve shows the mean value of the strength factors. }
    \label{fig: Ha, HeII, HeI s-factors wrt phase}
\end{figure}

The disentangling of the singly ionized $\mathrm{He \ II \ \lambda 4686}$ line is particularly interesting. Figure \ref{fig: disentangled HeII44686} shows the disentangled spectra of the line $\mathrm{He \ II \ \lambda 4686}$ in the low-hard (left) and the high-soft-intermediate (right) states, respectively. In the low-hard state, we identify a prominent absorption feature in the star's atmosphere and a weak blueshifted emission, which, however, may be due to an imprecision of the disentangled continuum. Yet, a completely different feature is observed in the focused wind. We observe a slightly blueshifted broad emission line in both the low-hard and high-soft-intermediate X-ray states. Nonetheless, in the low-hard state, the focused wind's emission line is more intense than in the high-soft state. \\
\indent From the disentangling method, we calculated the semi-amplitudes of the radial velocities for the hydrogen and helium lines, in the hard and soft-intermediate states for the stellar atmosphere component. We fixed the orbital parameters (e.g., orbital period, eccentricity and the periastron longitude) to the values found in the literature \citep{2007ragt.meet...71H}.
In the hard state versus the soft state, the semi-amplitude of the radial velocities of the different lines are: 79.1 $\mathrm{km \ s^{-1}}$ versus 76.9 $\mathrm{km \ s^{-1}}$ for He\,I $\mathrm{\lambda 6678}$ and 83.9 $\mathrm{km \ s^{-1}}$ versus 80.3 $\mathrm{km \ s^{-1}}$ for He\,II.

\subsection{Evolution of the strength factors with the orbital phase }

In addition to the orbital parameters, we calculated the strength factors $ s_j$. 
The top panel of Fig. \ref{fig: Ha, HeII, HeI s-factors wrt phase} shows the evolution of the strength factors with respect to the orbital phase, for the $\mathrm{H{\alpha}}$ (filled circles) and $\mathrm{He\ II}$ (cross) lines during the low-hard and high-soft-intermediate state, for the star's atmosphere component. We identified an orbital motion of the strength factor and fitted the data using the Fourier analysis:
\begin{equation}
    f(x) = a_0 + \sum_{n=1}^{2}  a_n \sin(2n\pi x + \phi_n)
.\end{equation}
The strength factors for $\mathrm{H{\alpha}}$ and $\mathrm{He\ II}$ are on average lower in the high-soft state than in the low-hard state. 
Furthermore, they reach two maxima at phase 0 and 0.5 -- at the conjunctions of the star -- for both the hydrogen and helium lines in both X-ray spectral states. Nevertheless, we can see that the strength factors at phase 0 are twice larger than the ones at phase 0.5. 
On the other hand, for He I they are on average constant with a mean value at 1 in both the hard and soft states (see the bottom panel of Fig. \ref{fig: Ha, HeII, HeI s-factors wrt phase}).

\subsection{The X-ray/optical anticorrelation}

\begin{figure*}[h]
    \centering
        \includegraphics[width=16cm]{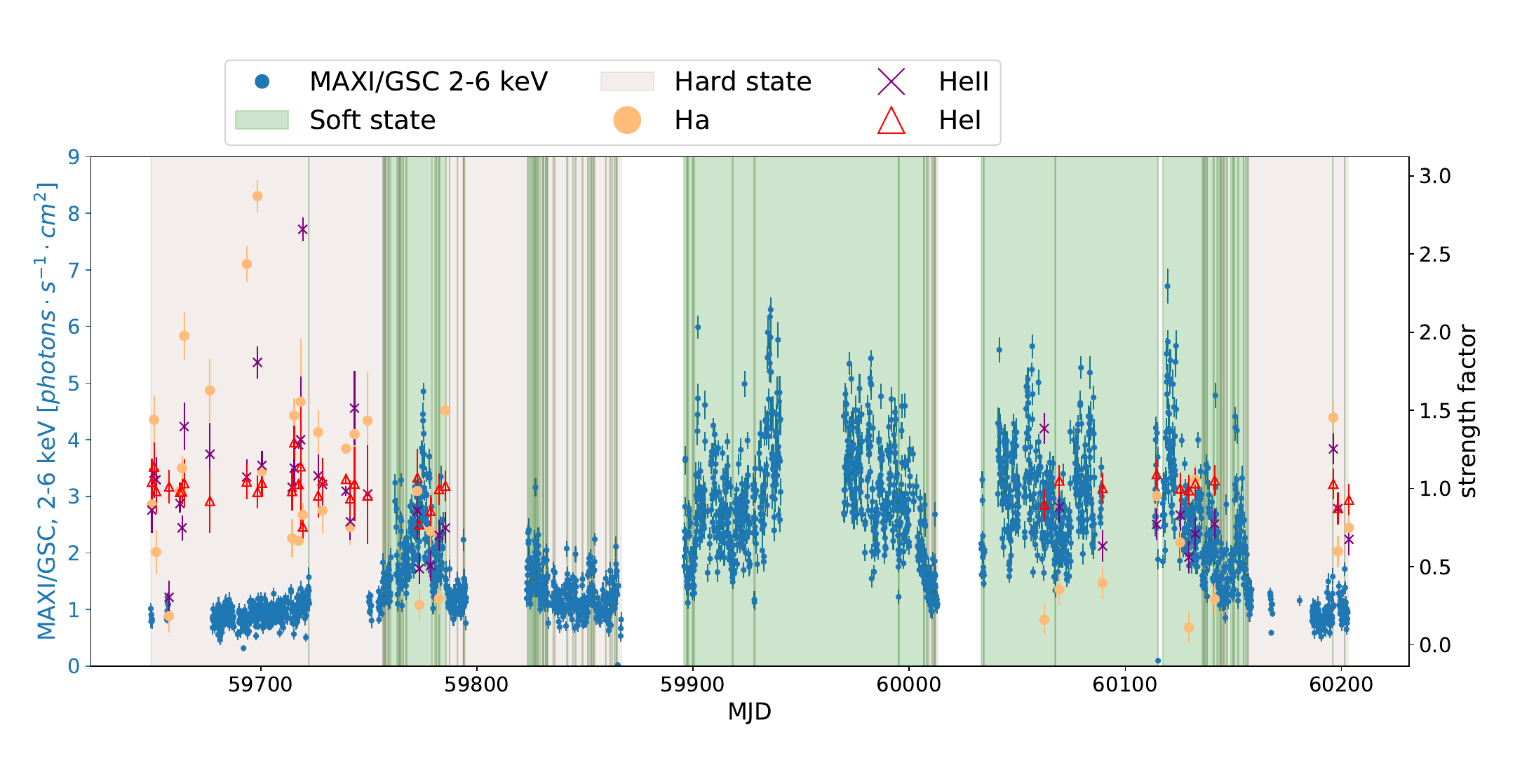}
     \caption{MAXI/GSC X-ray light curve in the 2--6 keV energy range (blue points) and optical strength factors from the stellar atmosphere component of the $\mathrm{H{\alpha}}$ (orange circles), He I (red triangles), and He II (purple cross) spectral lines during the entire period of the 2022--2023 observations.}
    \label{fig: s-factors star wrt X-rays}
\end{figure*}

X-ray variations in the accretion disk directly affect the ionization of the stellar wind, thereby influencing the optical emission. As a consequence, we would expect a direct correlation between the X-rays and the intensity of the optical emission. To analyze this correlation we consider the light curve from MAXI in the soft band (2--6 keV) and add the strength factors for the stellar wind component calculated for the hydrogen and helium spectral lines (see Fig. \ref{fig: s-factors star wrt X-rays}). The blue points represent the MAXI/GSC data. The orange circles, the red triangles and the purple crosses are the strength factors of $\mathrm{H{\alpha}}$, He I and He II, respectively. The shaded regions show the low-hard (shaded brown) and high-soft-intermediate (shaded green) X-ray spectral states. 

\noindent The disentangling of the combined spectra from both epochs in 2022 and in 2023 together show that the $s$-factors of the stellar and circumstellar components of H$\alpha$ and He\,II are in average higher in 2022 -- when the low-hard state prevails -- than in 2023 -- during the high-soft-intermediate state (see Fig. \ref{fig: s-factors star wrt X-rays} and the example of He II in Tables \ref{ondrejov2022} and \ref{ondrejov2023}). On the other hand, the absorption line of He\,I -- only observed in the spectrum of the stellar atmosphere -- remains practically unchanged (see Fig. \ref{fig: s-factors star wrt X-rays}). 

\begin{table}[h]
\caption{\label{table: anti-correlation coeffs} Results from the anticorrelation calculation for different components.}
\begin{threeparttable}[t]
    \begin{tabular}{rrrl}
\hline \hline 
a     & b      & $corr$      & component\\ [0.08cm]
\hline
1.759 & -0.503 & -0.496 & $\mathrm{H\alpha}$ stellar atmosphere\\[0.08cm]
1.918 & -0.609 & -0.667 & $\mathrm{H\alpha}$ CM\tnote{a} \ emission\\ [0.08cm]
1.305 & -0.202 & -0.407 & He\,II stellar atmosphere\\ [0.08cm]
1.185 & -0.122 & -0.236 & He\,II CM\tnote{a} \ emission\\[0.08cm]
0.984 &  0.010 &  0.067 & He\,I stellar absorption\\[0.08cm]
\hline
    \end{tabular}
    \tablefoot{
The coefficients $a$ and $b$ are from the anticorrelation relation between the s-factors and the X-ray flux: $s\simeq a+b F_X$, and $corr$ are the correlation parameters.\\
\tablefoottext{a}{Circumstellar matter, i.e, the focused wind.}
}
\end{threeparttable}
\end{table}

By interpolating the X-ray flux from MAXI/GSC data in the soft band (2--6 keV) on nights when our optical spectra were obtained, we find an (anti)correlation between the linear $s$-factors and the X-ray flux $\mathrm{F_X}$. The $s$-factors and the X-ray flux can be linked with the relation: $s\simeq a+b F_X$. For each spectral component, we calculated the values of the coefficients $a$, $b$ and the correlation parameter $corr$ given in Table \ref{table: anti-correlation coeffs}. If $corr = 1$ the $s$-factors and the X-ray flux are strongly correlated, for $corr = 0$ they are independent, and for $corr = -1$ they are anticorrelated. $\mathrm{H\alpha}$ and He II show higher values of $a$ and more negative values of $b$ than He I. Moreover, the correlation parameter $corr$ is negative for $\mathrm{H\alpha}$ and He II and an order of magnitude bigger than for He I. As a result, the $\mathrm{H{\alpha}}$ and He II lines show an anticorrelation between the strength factors and the X-ray flux.

\section{Discussion}\label{Discussion}
The above analysis shows that the spectral line profiles vary primarily due to orbital motion and changes of the X-ray spectral state, leading to changes in radial velocities and line intensities as traced by the strength factors. The disentangling enables the identification of the components contributing to this variability, which may have different origins.

\subsection{Variability in the optical photometry}

The optical photometric observations from LCO and TESS show that the variations in the optical light curves are not directly linked to the X-ray spectral states, but rather to stochastic and semi-regular aperiodic variations (see Figs. \ref{fig:LCO lc} and \ref{Fig:TESS}). These stochastic variations are consistent with the findings by \cite{2021A&A...648A..79K} who have shown that the stochastic photometric variability of OB stars are signatures of the instability caused by the line-driven winds. They argue that the photometric observational features could prove the existence of discrete overdensities embedded in large rarefied regions created by the line-driven wind instability. These so-called wind inhomogeneities (also known as "wind clumping" or "clumps") seem to already form close to the photosphere \citep[see also, e.g.,][]{2013A&A...559A}. The stochastic variations are also associated with the X-ray flares \citep{Karitskaya2000}. \citet{Petretti2021} analyzed TESS data from sector 14 (observational region of the sky) and attributed the variations to pulsations in the star. However, their irregular nature supposes that the changes in brightness are likely due to the physical processes happening in the stellar wind and the highly ionized surrounding material, where the variations in the magnitude occur at timescales shorter than the orbital period. \cite{1993A&A...279..485K} obtained similar results after analysis of the UV spectra of multiple HMXBs. Hence, detailed modeling of the data would be required to precisely explain the variability in the photometric light curves. \\

\subsection{Variability in the Balmer line profiles}

The monitoring of Cyg X-1 in the optical, in both the low-hard and high-soft-intermediate states, shows variations in the spectra of the Balmer lines, mainly the hydrogen and helium lines.
In Fig. \ref{fig: disentangled Ha} the redshifted emission line of the photosphere comes from the star photoionizing the wind in a receding region. The blueshifted absorption line shows photons re-scattering by the stellar wind, due to a net loss of photons in our line of sight. Compared to the photospheric emission feature, the focused wind emission is weaker because it does not necessarily propagate in the line of sight and is perturbed by the overall emission of the system. In the focused wind component, the blueshifted emission line is due to the photoionization from the star but also probably from the accretion disk because of their close proximity \citep[where the semi major axis of the system is estimated at 0.2 AU or $\mathrm{\approx 40 \ M_{\odot}}$; see][]{miller-jones_cygnus_2021}. The blueshifted emission also indicates a denser equatorial outflow, as previously suggested by \cite{2003ApJ...583..424G}. The weak, redshifted absorption is due to the ionized hot medium moving away from the line of sight. The inverse P-Cygni profile of the focused wind reveals that as part of the stellar atmosphere is ejected, some of it moves toward us in the blueshifted absorption of the wind, while the rest is accreted by the black hole through the focused wind. \\
\indent On the other hand, $\mathrm{He \ I \ \lambda 6678}$ only shows an absorption feature in the stellar atmosphere component and is not detected in the focused wind. The absence of emission indicates that the helium line is found in the photospheric region of the star. \\
$\mathrm{\indent He \ II \ \lambda 4686}$ is a more puzzling case to analyze because of the absorption detected in the stellar atmosphere and the emission detected in the focused wind. 
One possible explanation for the broad emission line is that the faster stellar wind and the slower (denser) matter of the focused wind meet and interact in the region closer to the black hole, because of the wind stagnation induced by the high X-ray radiation. As a result, many shocks and inhomogeneities (in both density and velocity) can form and produce the emission lines.
Strong stellar winds from O-B stars eject matter in the interstellar medium with high velocities \citep[with an approximate terminal velocity of $\mathrm{2100 \ km \ s^{-1}}$ for HDE 226868; see][]{1995A&A...297..556H}. In the low-hard state, when the X-ray emission decreases, the wind is less destroyed by the X-ray irradiation and the wind's inhomogeneities are more important. The clumps are denser, bigger and scatter more than in the high-soft state, leading to a higher intensity in emission. Nevertheless, a proper modeling of the $\mathrm{He \ II \ \lambda 4686}$ line is necessary to constrain the properties of the medium in which the line forms and propagates, as well as the clumping properties leading to the high re-scattering.\\

\subsection{Variations in the radial velocity curves}

The radial velocity measurements using the method of Fourier disentangling indicate a maximum radial velocity of the stellar atmosphere at 89.7 $\mathrm{km \ s^{-1}}$ in absolute value, which is consistent with the $\mathrm{H{\alpha}}$ equivalent width measurements from \cite{2003ApJ...583..424G} and \cite{2008ApJ...678.1237G}, respectively at 81.3 $\mathrm{km \ s^{-1}}$ and 88.6 $\mathrm{km \ s^{-1}}$.
Furthermore, we observe an antiphase motion between the star and the focused wind traced respectively by He I and He II in Fig. \ref{fig: RV from HeI6678 and HeII4686}. The antiphase motion between the two components shows that while the stellar wind expands, the focused wind transfers matter away toward the black hole, with an average delay in phase of 0.25, so approximately 1.5 days. This delay also corresponds to the time-period variability between a minimum and a maximum brightness observed in the photometric data from TESS (see Fig. \ref{Fig:TESS}). This antiphase motion has also been observed in previous studies \citep{1987ApJ...321..438N, 2008AJ....136..631Y}. It indicates a focused wind geometry elongated toward the black hole and a conical emission from the focused wind in the region facing the black hole. The region facing the black hole is more ionized than the region on the other side of the star. Thus, the semi-amplitude of the velocities will change more dramatically in the region facing the black hole than in the hidden side of the star. The anti-orbital motion and the emission line profile of He II confirm a higher density in the focused wind region, which matches the results from \cite{2008ApJ...678.1237G} and \cite{2015A&A...575A...5C}, suggesting a denser outflow in the equatorial plane. \\

\subsection{Variability due to the X-ray emission}

The line profile also varies depending on the spectral state. 
Variations between the low-hard and high-soft-intermediate states result from changes in X-ray heating, which in turn affect the ionization state of the wind.
The spectral lines are more intense in the low-hard state, which has also been observed in \cite{2008ApJ...678.1237G}. In the soft state, the X-ray flux is more intense, the ionization state of the wind increases and more ions are losing bound electrons. The wind being line-driven, it needs to have enough ions with bound electrons to be able to absorb photons in the optical/UV range, which increases the radiation pressure and the velocity of the wind.
Therefore, the wind is faster when the X-ray flux decreases in the low-hard state. \cite{2008ApJ...678.1237G} found similar results using  UV spectra of Cyg X-1 from the Space Telescope Imaging Spectrograph (STIS) from the Hubble Space Telescope. This can also be seen in the calculation of the radial velocities' semi-amplitude with an increase of 6 $\mathrm{km \ s^{-1}}$ in the low-hard state when the stellar wind is the strongest. \\
\indent In the low-hard state, the X-ray flux fluctuates between 0.8 and 2.6 $\mathrm{photons \ s^{-1} \ cm^{-2}}$ (see Table \ref{ondrejov2023}). In this state, the wind is less ionized, and thus the opacity and the radiative pressure increase. The ionization parameter is proportional to the X-ray luminosity with the relation
\begin{equation}
    \xi = \frac{L_X}{n_p \ r^2} 
,\end{equation}
with ${n_p}$ the proton number of density and $r$ the radial distance to the X-ray source. 
As a consequence, when the wind's density increases, for a constant distance with respect to the X-ray source, the ionization parameter decreases. Less ionization means a smaller ionization region, a stronger wind, and the $\mathrm{H{\alpha}}$ and He II emission and absorption lines' intensity increases. The distance between the ionizing source and the absorbing region is estimated to be around $\mathrm{3 \times 10^{11} - 10^{12}}$ cm (0.02 AU) by \cite{2005ApJ...620..398M}.
Moreover, a faster wind means broader line profiles in the low-hard state due to Doppler broadening, which depends on the velocity distribution of the wind: the faster the wind, the broader the emission lines. However, for the focused wind component for $\mathrm{H{\alpha}}$, we observed a broader line in the high-soft-intermediate state than in the low-hard state. As a result, the focused wind moves faster in the high-soft-intermediate state, where more matter is accreted by the black hole. \\

\subsection{Variations due to the orbital motion}

However, the line profile also depends on the orbital motion of the system. During the ejecta of matter, the wind faces strong density and temperature inhomogeneities. 
These wind inhomogeneities commonly referred to as "clumps" can already form in the photosphere. Previous studies support optically thick clumps \citep{2010A&A...521L..55P, 2008cihw.conf.....H}, through spectroscopic measurements of the radial optical depths ratios of the red and blue counterparts of the Si IV doublet of B supergiants. Furthermore, the denser regions containing optically thick clumps show more prominent absorption and emission features in the spectra \citep[also observed in the X-ray spectra from][]{2023A&A...680A..72H}. Therefore, at the inferior conjunction -- when the star is in front of the black hole along the line of sight -- the wind is the strongest and the most absorbed by the clumps, increasing the s-factors at phase 0 (see Fig.\ref{fig: Ha, HeII, HeI s-factors wrt phase}). On the opposite side -- at the superior conjunction -- the star is further away, and the wind is weaker, so the lines exhibit a dominant P-Cygni profile and are less absorbed. These results confirm the larger emission features at the conjunctions observed by \cite{2003ApJ...583..424G}.\\ 
\indent The two maxima of the s-factors at phases 0 and 0.5, for both the hydrogen and helium lines, in both X-ray spectral states, indicate a higher dependence on the orbital motion than the spectral state. Moreover, the s-factors are higher in the low-hard state, which confirms that the wind has deeper absorption lines and higher emission lines in the hard state compared to the high-soft-intermediate state. \\

\subsection{X-ray/optical anticorrelation}

These results show that the X-ray irradiation from the accretion disk has a direct impact on the stellar wind motion and the optical emission. Inversely, the stellar wind has an impact on the accretion disk and probably also on the accretion rate of the black hole. Intuitively, we expect the black hole accretion rate to increase when the mass loss of the star increases. Nevertheless, in the Bondi accretion approximation \citep{1944MNRAS.104..273B, 1976A&A....49..327L} the accretion rate is also directly related to the wind velocity via the relation
\begin{equation}
    \dot{m}_{accr} \propto \Dot{M} v^{-4}
,\end{equation}

\noindent with $\Dot{M} = \frac{L \tau}{v_{\infty} c}$ being the mass loss rate of the star, $L$ is the stellar luminosity, $\tau$ is the optical depth of the wind and $v_{\infty}$ is the terminal velocity of the wind. Thus, a fast line-driven wind should generate a lower accretion rate, even for a high stellar mass loss. \\
\indent Previous studies showed that the terminal velocity of the wind is strongly dependent on the X-ray spectral state. In the hard state, \cite{1983ApJ...270..671D} find $v_{\infty} = 2300 \ \mathrm{km \ s^{-1}}$ and \cite{1995A&A...297..556H} find $v_{\infty} = 2100 \ \mathrm{km \ s^{-1}}$. Whereas in the soft state, \cite{2008ApJ...678.1248V} find $v_{\infty} = 1420 \ \mathrm{km \ s^{-1}}$ and \cite{2008ApJ...678.1237G} find $v_{\infty} = 1200 \ \mathrm{km \ s^{-1}}$, with a real value estimated closer to $v_{\infty} = 1600 \ \mathrm{km \ s^{-1}}$. 
These results indicate that the wind's terminal velocity is higher in the low-hard state, when the wind absorption dominates, confirming a faster wind in the hard state based on our optical analysis. The mass loss rate of the star is also dependent on the X-ray spectral state and the clumping properties of the wind \citep[see][]{2018A&A...620A.150K}. \cite{2003ApJ...583..424G} estimated the mass loss rate to be $\mathrm{\Dot{M}_{hard} = 2.57 \ \cdot \ 10^{-6} \ M_{\odot} \ yr^{-1}}$ in the hard state versus $\mathrm{\Dot{M}_{soft} = 2.00 \ \cdot \ 10^{-6} \ M_{\odot} \ yr^{-1}}$ in the soft state. \cite{Wen_1999} similarly observed a higher mass loss rate in the low-hard state based on the analysis of the X-ray orbital light curves. By considering that the wind velocity is equal to the terminal velocity, we can roughly estimate the accretion rate using the previous observational data:
\begin{equation}
    \frac{\Dot{m}_{accr, hard}}{\Dot{m}_{accr, soft}} = 
    \frac{\Dot{M}_{hard}}{\Dot{M}_{soft}} \frac{v_{hard}^{-4}}{v_{soft}^{-4}} \approx \ 0.3
.\end{equation}
This confirms the observational results in X-rays where the accretion rate in the soft state is higher than in the hard state. This shows that when the accretion disk's irradiation increases, the wind density drops, and the line-driven wind decelerates -- as well as the focused wind -- which allows more matter to be accreted by the black hole and the accretion rate should increase. Consequently, changes in the wind velocity influence the accretion rate, which is important in our understanding on the origin of the change of states for X-ray binaries. Thus, the stellar wind should influence the accretion rate of the accretion disk and hence, contribute to the X-ray spectral state transitions in HMXBs. Nevertheless, a deeper analysis is required to identify how much the stellar wind is contributing to changes in the accretion rate in addition to the accretion disk instabilities.

\section{Conclusions}\label{discussion and ccl}

 We performed high-resolution spectroscopy in the hard and soft-intermediate states of the HMXB Cyg X-1 and applied the method of Fourier disentangling to separate the components coming from the stellar atmosphere and the focused wind. First, we show that $\mathrm{H\alpha}$ exhibits a P-Cygni profile in the stellar atmosphere component, meaning that this component is part of the outflowing material from the star. The He I $\mathrm{\lambda 6678}$ absorption line is only observed in the stellar atmosphere and confirms an origin in the photospheric region of the star. On the other hand, the He II $\mathrm{\lambda 4686}$ line in absorption in the stellar atmosphere and in emission in the focused wind seems to come from collisions or recombination processes in the higher-density gradients of the focused wind, due to strong inhomogeneities in the wind. One remaining question would be what the structure of these clumps is and how much they contribute to the overall absorption of the optical emission. 

Second, the stellar wind is stronger and faster in the low-hard spectral state because of the low ionization of the line-driven wind. Nevertheless, the line profile also depends on the orbital motion of the system, where the line intensity also dominates at the inferior conjunction of the star because of stronger clump absorptions in the line of sight. In a follow-up analysis, we plan to apply the method of Fourier disentangling to the UV spectra to perform a deeper analysis of the wind in both states simultaneously with optical spectra. Using UV spectra is advantageous when trying to precisely determine the wind’s parameters in different X-ray accretion states.\\
 \indent Ultimately, the anticorrelation between the X-ray flux (and thus the X-ray spectral state) and the optical emission from the companion star suggests a strong interplay between the stellar wind and the accretion rate of the black hole. Studying the time delay between the matter being ejected and accreted can provide valuable insights into how the matter is being accreted by the black hole and the physical processes governing the accretion process in HMXBs. 

\section*{Data availability}
The results from the \texttt{KOREL} code for the disentangling method can be found in the following link:
\url{https://doi.org/10.5281/zenodo.14800227}.

\begin{acknowledgements}
        Research is based on data taken with the Perek telescope at the Astronomical Institute of the Czech Academy of Sciences (ASU) in Ond\v{r}ejov. This material is also based upon work supported by Tamkeen under the NYU Abu Dhabi Research Institute grant CASS.  
        M.B. acknowledges the support from GAUK project No. 102323. JS thanks GACR project 21-06825X for the support. B.K. and M.C. acknowledge the support from the Grant Agency of the Czech Republic (GA\v{C}R 22-34467S) and RVO:67985815. 
        We thank the Stellar Physics Department of ASU and their technical staff for the support during the observations. We thank Daniel Bramich for the “X-ray Binary New Early Warning System (XB-NEWS)” pipeline and the LCO data reduction, Anastasiya Yilmaz for the artistic representation of Cyg X-1 in Figure \ref{fig:sketch anastasiya}, and Victoria Grinberg for valuable discussions. 
    
\end{acknowledgements}

%
   \bibliographystyle{aa} 
   \bibliography{ref} 

%


\begin{appendix}
\onecolumn   

\section{Additional disentangled spectra}

\begin{figure*}[h!]
    \centering
    \includegraphics[width=6cm]{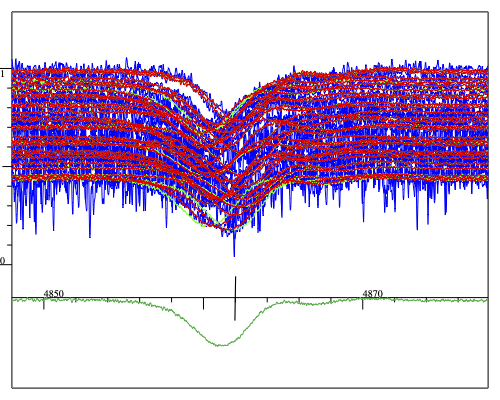}
    \includegraphics[width=6cm]{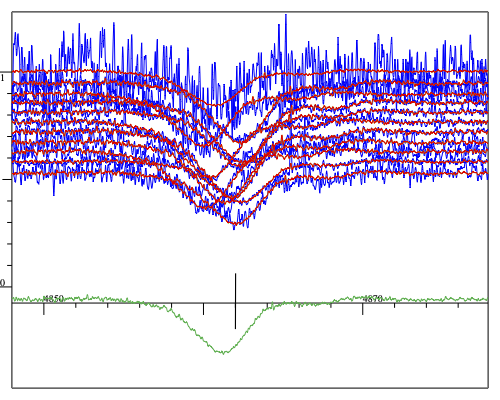}
    \caption{Disentangled spectra of $\mathrm{H{\beta}}$ with a central wavelength at $\mathrm{4860 \ \AA}$, indicated with the vertical black line. \textit{Left}: Disentangled spectra from the 2022 epoch in the hard X-ray spectral state. \textit{Right}: Disentangled spectra from the 2023 epoch in the soft X-ray spectral state. The top of each plot combines the OES observed spectra (blue) and the fit by the separated spectra of the components (red) at different orbital phases. The green line shows the disentangled line from the star's atmosphere. We do not detect the focused wind spectral component.}
    \label{fig: disentangled Hb}
\end{figure*}


\begin{figure*}[h!]
    \centering
    \includegraphics[width=6cm]{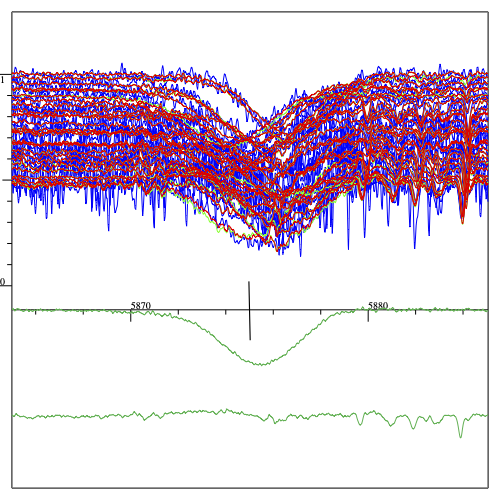}
    \includegraphics[width=6cm]{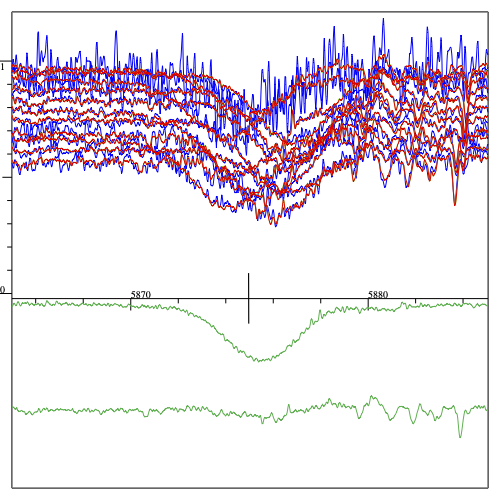}
    \caption{Disentangled spectra of He I with a central wavelength at $\mathrm{5875 \ \AA}$, indicated with the vertical black line. The green lines are the disentangled lines from the star's atmosphere and the telluric lines.}
    \label{fig: disentangled HeI5875}
\end{figure*}

\section{Observation logs of the Ondřejov optical spectra}

\begin{table*} [h]     
    \caption{\label{ondrejov2022}Observation log of the Ondřejov optical spectra taken in 2022, when the hard state lasted the longest.}
\centering                          
\begin{tabular}{c c c c c c c}        
\hline\hline                 
 Date                                   &       HJD             & Exposure time   & Orbital phase $\mathrm{\phi}$ & $\mathrm{s_{star}}$ for He II & $\mathrm{F_{X}}$ & Instrument  \\  [0.08cm]  
 [UT]                                   &                               &               [ks]    &                                                       &                               &       $\mathrm{[photons \ s^{-1} \ cm^{-2}]}$   &               \\ [0.08cm]
 \hline  \\[-1.8ex]
 2022-03-11 02:23:10.666        &  2459649.60   &       3.6     &       0.178         &       0.9086  &       0.8617          &   OES \\      [0.08cm]
 2022-03-12 02:24:08.640        &  2459650.60   &       3.4     &       0.356         &       1.1581  &       0.8726          &   OES \\      [0.08cm]
 2022-03-13 02:48:32.688        &  2459651.62   &       3.6     &       0.538         &       1.1086  &       0.9100          &   OES \\      [0.08cm]
 2022-03-19 02:02:13.286        &  2459657.59   &       3.6     &       0.604         &       0.2745  &       1.0303          &   OES \\      [0.08cm]
 2022-03-24 02:17:02.429        &  2459662.60   &       3.6     &       0.499         &       0.9660  &       0.9912          &   OES \\      [0.08cm]
 2022-03-24 03:22:50.563        &  2459662.64   &   3.0 &   0.507       &                       &                               &  CCD700 \\      [0.08cm]        
 2022-03-25 02:19:31.210        &  2459663.60   &       3.6     &       0.677         &       0.7103  &       0.9844          &   OES \\  [0.08cm]
 2022-03-26 01:52:07.968        &  2459664.58   &       3.6     &       0.852         &       1.3852  &       0.9771          &   OES \\  [0.08cm]
 2022-03-26 03:07:38.179    &  2459664.63   &   3.9     &   0.862       &                       &                               & CCD700  \\      [0.08cm]        
 2022-04-06 23:37:04.598        &  2459676.49   &       3.6     &       0.979         &       1.2042  &       0.8471          &   OES \\  [0.08cm]
 2022-04-07 00:43:33.773    &  2459676.53   &   3.0 &   0.987   &                       &                               & CCD700  \\      [0.08cm]
 2022-04-23 23:52:07.478        &  2459693.50   &       3.6     &       0.016         &       1.0740  &       0.8212          &   OES \\  [0.08cm]
 2022-04-24 00:58:46.675    &  2459693.54       &       3.0     &   0.025         &                       &                               & CCD700        \\      [0.08cm]
 2022-04-29 00:00:33.091        &  2459698.50   &       3.6     &       0.910         &       1.7754  &       0.9424          &   OES \\  [0.08cm]
 2022-04-29 01:05:45.283        &  2459698.55   &       3.6 &   0.918   &                       &                               & CCD700  \\      [0.08cm]
 2022-05-01 01:09:20.246        &  2459700.55   &       3.6     &       0.276         &       1.2073  &   0.9538              &   OES \\      [0.08cm]
 2022-05-13 00:02:33.446        &  2459712.50   &       3.9     &   0.411         &                       &                               &       CCD700  \\     [0.08cm]
 2022-05-14 22:50:08.995        &  2459714.45   &       3.6     &   0.759         &                       &                               &       CCD700  \\     [0.08cm]
 2022-05-15 00:00:42.250        &  2459714.50   &       4.0     &       0.767         &       0.9778  &       1.0034          &   OES \\      [0.08cm]
 2022-05-15 22:28:25.824    &  2459715.44   &   3.6     &   0.935       &                       &                               & CCD700  \\      [0.08cm]
 2022-05-15 23:38:19.075        &  2459715.49   &       4.0     &       0.943         &       1.1321  &       0.9183          &   OES \\  [0.08cm]
 2022-05-18 00:45:00.000        &  2459717.53   &       3.6     &       0.309         &       1.3393  &       1.0157          &   OES \\  [0.08cm]
 2022-05-18 22:26:09.398        &  2459718.44   &       4.0     &       0.470         &       1.3948  &       1.0854          &   OES \\  [0.08cm]
 2022-05-18 23:46:23.693        &  2459718.49   &   4.0 &   0.480       &                       &                               & CCD700  \\      [0.08cm]
 2022-05-19 22:09:41.155        &  2459719.43   &       2.5     &       0.647         &       2.0245  &       1.1132          &   OES \\  [0.08cm]
 2022-05-26 23:10:28.358        &  2459726.47   &       3.6     &       0.904         &                       &                               &  CCD700       \\      [0.08cm]
 2022-05-27 00:49:26.630        &  2459726.54   &       3.6     &       0.917         &       1.0351  &       1.1339          &   OES \\      [0.08cm]
 2022-05-29 01:17:23.914        &  2459728.56   &       3.6     &       0.277         &       1.0850  &       1.1578          &   OES \\      [0.08cm]
 2022-06-08 22:25:12.115        &  2459739.44   &       4.0     &       0.220         &       1.0454  &       1.3631          &   OES \\      [0.08cm]
 2022-06-08 23:44:40.272        &  2459739.49   &   4.0 &   0.230       &                       &                               &  CCD700 \\      [0.08cm]
 2022-06-10 21:54:27.130        &  2459741.42   &       4.0     &       0.574         &       0.7523  &       1.3839          &   OES \\      [0.08cm]
 2022-06-12 21:15:01.930        &  2459743.39   &       3.6     &       0.926         &       1.4870  &       1.3876          &   OES \\      [0.08cm]
 2022-06-12 22:30:09.158    &  2459743.44       &       3.6     &       0.935         &                       &                               & CCD700  \\     [0.08cm]
 2022-06-18 22:50:07.699        &  2459749.45   &       3.6     &       0.009         &       0.9884  &       1.1555          &   OES \\      [0.08cm]
 2022-06-19 00:04:29.827        &  2459749.51   &       3.6     &   0.018         &                       &                               & CCD700        \\      [0.08cm]
 2022-07-11 23:45:20.016        &  2459772.49   &       3.6     &       0.123         &       0.8856  &       2.4229          &   OES \\      [0.08cm]
 2022-07-12 20:30:00.720        &  2459773.36   &       3.6     &       0.277         &                       &                               & CCD700        \\      [0.08cm]
 2022-07-12 21:34:45.782        &  2459773.40   &       3.6     &       0.285         &       0.5233  &       2.6149          &   OES \\      [0.08cm]
 2022-07-17 20:39:56.534        &  2459778.36   &       3.6 &   0.172   &                       &                               & CCD700  \\      [0.08cm]
 2022-07-18 01:05:19.882        &  2459778.55   &       2.7     &       0.204         &       0.5409  &       2.1481          &   OES \\      [0.08cm]
 2022-07-21 22:50:35.261        &  2459782.45   &       3.6 &   0.902   &                       &                               & CCD700  \\      [0.08cm]
 2022-07-21 23:56:17.347        &  2459782.50   &       3.6     &       0.910         &       0.7003  &       1.7805          &   OES \\      [0.08cm]
 2022-07-24 20:32:51.533        &  2459785.36   &       3.6 &   0.421   &                       &                               &  CCD700 \\      [0.08cm]
 2022-07-24 22:13:37.632        &  2459785.43   &       3.6     &       0.433         &       0.7921  &       1.5820          &   OES \\  [0.08cm]
\hline  
\end{tabular}
\end{table*}

\begin{table*}[h]
\centering   
   \caption{\label{ondrejov2023}Same as Table~\ref{ondrejov2022} but for 2023, when the soft-intermediate state lasted the longest.}
\begin{tabular}{c c c c c c c}             
\hline\hline       
\rule{0cm}{0.3cm}
 Date                                   &       HJD             & Exposure time   & Orbital phase $\mathrm{\phi}$ & $\mathrm{s_{star}}$ for He II & $\mathrm{F_{X}}$ & Instrument  \\   [0.08cm] 
 [UT]                                   &                               &               [ks]    &                                                       &                               &       $\mathrm{[photons \ s^{-1} \ cm^{-2}]}$   &       \\ [0.08cm]
 \hline  \\[-1.8ex]
2023-04-28 00:25:23.923         &       2460062.52      &       3.6             &               0.915   &       1.4166  &       2.6443  &               OES                     \\ [0.08cm]
2023-04-28 01:39:40.176         &       2460062.57      &       3.6             &               0.925         &                       &                       &               CCD700          \\ [0.08cm]
2023-05-05 00:13:28.618         &       2460069.51      &       3.6             &               0.164         &       0.9450  &       2.3452  &               OES                     \\ [0.08cm]
2023-05-05 01:00:18.778         &       2460069.54      &       3.6             &               0.170         &                       &                       &               CCD700          \\ [0.08cm]
2023-05-25 00:42:36.144         &       2460089.53      &       3.6             &               0.740         &       0.6029  &       3.0157  &               OES                     \\ [0.08cm]
2023-05-25 01:45:36.317         &       2460089.58      &       1.8             &               0.747         &                       &                       &               CCD700          \\      [0.08cm]
2023-06-18 23:16:06.787         &       2460114.47      &       3.6             &               0.193         &                       &                       &               CCD700          \\      [0.08cm]
2023-06-19 00:20:54.960         &       2460114.52      &       3.6             &               0.201         &       0.8110  &       3.0874  &               OES                     \\  [0.08cm]
2023-06-29 20:33:27.302         &       2460125.36      &       3.6             &               0.137         &                       &                       &               CCD700          \\      [0.08cm]                
2023-06-29 21:39:06.451         &       2460125.40      &       3.6             &               0.145         &       0.8851  &       2.6557  &               OES                     \\  [0.08cm]
2023-07-03 21:50:18.557         &       2460129.41      &       3.6             &               0.861         &                       &                       &               CCD700          \\      [0.08cm]        
2023-07-03 22:54:26.726         &       2460129.45      &       3.6             &               0.869         &       0.5505  &       2.4501  &               OES                     \\  [0.08cm]
2023-07-04 01:23:38.976         &       2460129.56      &       2.6             &               0.888         &                       &                       &               CCD700          \\      [0.08cm]
2023-07-06 20:59:13.085         &       2460132.38      &       4.2             &               0.391         &                       &                       &               CCD700          \\      [0.08cm]
2023-07-06 22:13:53.184         &       2460132.43      &       4.2             &               0.400         &       0.7407  &       2.3463  &               OES                     \\      [0.08cm]
2023-07-15 20:25:37.805         &       2460141.35      &       3.6             &               0.993         &       0.7209  &       2.1947  &               OES                     \\      [0.08cm]
2023-07-15 21:42:25.862         &       2460141.41      &       3.6             &               0.003         &                       &                       &               CCD700          \\      [0.08cm]
2023-09-08 18:37:35.472         &       2460196.28      &       3.6             &               0.801         &       1.2337  &       1.0473  &               OES                     \\  [0.08cm]
2023-09-08 19:52:22.310         &       2460196.33      &       3.6             &               0.811         &                       &                       &               CCD700          \\      [0.08cm]
2023-09-10 19:12:33.869         &       2460198.30      &       3.6             &               0.163         &                       &                       &               CCD700          \\      [0.08cm]
2023-09-10 20:57:30.614         &       2460198.38      &       3.6             &               0.176         &       0.9174  &       1.1046  &               OES                     \\  [0.08cm]
2023-09-15 23:13:17.270         &       2460203.47      &       3.6             &               0.086         &                       &                       &               CCD700          \\      [0.08cm]
2023-09-16 00:31:47.107         &       2460203.52      &       3.6             &               0.095   &       0.7002  &       0.9783  &               OES                     \\      [0.08cm]
\hline                                                                                                  
                               
\end{tabular}
\tablefoot{The orbital phases are calculated from the ephemeris defined at the inferior conjunction of the companion star from \cite[]{2003ApJ...583..424G}: $2 451 730.449(8) + 5.599829(16)\times E$. The X-ray flux $\mathrm{F_X}$ is taken from MAXI/GSC data.}
\end{table*}

\end{appendix}

\end{document}